\documentclass[a4paper,fleqn,usenatbib,useAMS]{mnras}

\usepackage{graphicx}	
\usepackage{amsmath}	
\usepackage{amssymb}	
\usepackage{multicol}        
\usepackage{bm}		
\usepackage{pdflscape}	

\newcommand{\bibtex}{\textsc{Bib}\!\TeX} 

\usepackage[T1]{fontenc}
\usepackage{ae,aecompl}

\usepackage{newtxtext,newtxmath}

\title[Dust attenuation  curve of z$\sim$2 (U)LIRGs
]{Characterizing the UV-to-NIR shape of the dust attenuation curve of IR luminous galaxies up to z$\sim$2
}

\author[B.~Lo Faro et al.]{B. Lo Faro$^{1}$, V. Buat$^{1}$\thanks{e-mail: veronique.buat@lam.fr}, Y. Roehlly$^{1,2}$, J. Alvarez-Marquez$^{1,3}$, D. Burgarella$^{1}$,
\newauthor L. Silva$^{4}$, A. Efstathiou$^{5}$\\
$^{1}$Aix Marseille Univ, CNRS, LAM, Laboratoire d'Astrophysique de Marseille, Marseille, France\\
$^{2}$Astronomy Centre, Department of Physics and Astronomy, University of Sussex, Brighton BN1 9QH\\
$^{3}$Departamento de astrofísica, Centro de Astrobiología (CAB, CSIC-INTA), Carretera de Ajalvir, 28850 Torrejón de Ardoz, Madrid, Spain\\
$^{4}$National Institute for Astrophysics INAF-OATs, Trieste, Italy\\
$^{5}$School of Sciences, European University Cyprus, Diogenes Street, Engomi, 1516 Nicosia, Cyprus}

\date{Last updated 2015 May 22; in original form 2013 September 5}

\pubyear{2015}

\begin{document}
\label{firstpage}
\pagerange{\pageref{firstpage}--\pageref{lastpage}}
\maketitle

\begin{abstract}
In this work we investigate the far-UV to NIR shape of the dust attenuation curve of a sample of IR selected dust obscured (U)LIRGs at z$\sim$2.
The spectral energy distributions (SEDs) are fitted with CIGALE, a physically-motivated spectral synthesis model based on energy balance. Its flexibility allows us to test a wide range of different analytical prescriptions for the dust attenuation curve, including the well-known Calzetti and Charlot \& Fall curves, and modified versions of them.
The attenuation curves computed under the assumption of our reference double power-law model  are in very good agreement with those derived, in previous works, with radiative transfer (RT) SED fitting. We investigate the position of our galaxies in the IRX-$\beta$ diagram and find this to be consistent with grayer slopes, on average, in the UV. We also find evidence for a flattening of the attenuation curve  in the NIR with respect to more classical Calzetti-like recipes. This larger NIR attenuation yields larger derived stellar masses from SED fitting, by a median factor of $\sim$ 1.4 and up to a factor $\sim$10 for the most extreme cases. The star formation rate appears instead to be more dependent on the total amount of attenuation in the galaxy. Our analysis highlights the need for a flexible attenuation curve when reproducing the physical properties  of a large variety of objects.
\end{abstract}

\begin{keywords}
galaxies: evolution -- galaxies: general -- galaxies: spectral synthesis -- galaxies: IR emission -- galaxies: dust attenuation -- infrared: galaxies
\end{keywords}




\section{Introduction}

Dust plays a crucial role in many aspects of galaxy formation and evolution, in particular by strongly affecting the spectral energy distribution (SED): it absorbs and scatters photons, mostly at wavelengths $< 1 \mu$m, and thermally emits the absorbed energy in the Infrared (IR) ($\lambda \sim 1-1000 \mu$m). The ratio of the IR to UV luminosity, usually referred as IR excess, (IRX=$Log(L_{\rm IR}/L_{\rm UV})$), is tightly related to the amount of dust attenuation in a galaxy. Since the original work of \citet{Meurer1999} on local UV-bright starburst galaxies, the relation between the IR excess and the UV continuum slope, $\beta$, has been intensively used at all redshift (particularly at high-$z$) to estimate the total amount of attenuation in the UV.
However, important deviations from this relation have been observed both for galaxies forming stars at a lower rate, as commonly found in the nearby universe \citep[see][]{Buat2002, Buat2005, Gordon2004, Kong2004, Calzetti2005, Seibert2005, Boissier2007, Dale2009, Boquien2009, Boquien2012}, and for IR bright galaxies at all redshifts \citep[][]{Howell2010, Reddy2012, Casey2014}. The interpretation of the IRX-$\beta$ relation and of its deviations appears to require either different attenuation laws or star formation histories (or both) \citep[see e.g.][]{Kong2004, Panuzzo2007, Conroy2010, Forrest2016, Salmon2015}.

The dust attenuation curve of a galaxy is the result of the complex blending of both the optical and physical properties of dust and the relative geometrical distribution of stars and dust within a galaxy \footnote{Having to deal with geometrical effects, estimating dust attenuation is different than estimating dust extinction. With the term \textit{extinction} we refer to the wavelength dependence of the optical properties (absorption plus scattering) of the dust mixture. These are measured in the simple geometrical configuration of a homogeneous slab of dust placed between the observer and a point source as in the case when directly measuring the extinction from observations of background stars. In this configuration a direct relation between the ratio of the observed to intrinsic luminosity and the dust optical depth can be easily defined \citep[see][and references therein for more details]{Conroy2013}}.
All these complex aspects are usually accounted for in radiative transfer models with varying levels of sophistication, with a general indication that shallower attenuation curves
can be ascribed to larger optical depths and mixed star-dust geometries possibly masking variations of the intrinsic dust properties \citep[e.g.][]{CharlotFall00, Granato2000, WittGordon2000, Calzetti2001, Pierini2004, Tuffs2004, Panuzzo2007, Chevallard2013}.
On the observational side most of the works dealing with dust attenuation estimates in galaxies are usually tied to UV wavelengths where dust effects are dominant and where most of the young massive stars dominating the current star formation rate emit their photons. In their seminal work, \citet{Calzetti1994} estimated a mean attenuation curve from a sample of 39 local UV-bright starburst using UV-optical spectra. Under the assumption of the same underlying stellar population for their galaxies, they were able to build average attenuation curves by comparing galaxies with high and low Balmer line ratios, i.e., by comparing dusty to dust-free galaxies \citep[][]{Calzetti2001, Conroy2013}. Their measured attenuation curve is characterized by a grayer slope than both Milky Way  and Large Magellanic Cloud  extinction curves and by the  lack of the 2175 \AA~absorption feature.

 Recently \citet{Battisti2016, Battisti2017}  combined UV photometric data from GALEX with SDSS spectroscopy and near-IR data from the UKIRT and 2MASS surveys  for several thousands of local galaxies and implemented a method close to the one used by \citet{Calzetti1994}. They derived an attenuation curve  similar to the one of \citet{Calzetti2000} although slightly lower in the UV.
\citet{Reddy2015} used a similar method at high redshift to estimate the attenuation curve of z$\sim$2 galaxies selected from the MOSDEF survey and found the shape to be similar to the Calzetti one in the UV and steeper at longer wavelength. More recently \citet{Reddy2016} presented the first measurements of the shape of the far-ultraviolet (950 \AA~$<\lambda<$ 1500 \AA~) dust attenuation curve at z$\sim$3 characterized by a lower attenuation in the far-UV for a given E(B-V) than standard recipes \citep[][]{Calzetti2000, Reddy2015}.

Important insights into the characterization of the dust attenuation curve in different galaxy samples have been also provided by SED-fitting models \citep[][]{Ilbert2009, Conroy2010, Buat2010, Buat2011, Buat2012, Wild2011b, KriekConroy2013, Salmon2015}. These codes  optimized either to measure photometric redhifts or to derive physical parameters like star formation rate (SFR) or stellar mass (M$_{\star}$)  assume an attenuation law and most of them explore different scenarios and put constrains on them. The main advantage of these studies is that they allow to consider very large samples of galaxies and/or to search for  variations of the attenuation law as a function of different quantities. Variations of the UV ``steepness'' of the dust attenuation curve have been observed for strong UV emitting galaxies \citep[see e.g.,][]{Conroy2010, Buat2011, Buat2012, KriekConroy2013}, as a function of the position in the IRX-$\beta$ plane and/or color excess \citep[e.g.,][]{Salmon2015}. Variations of the shape of the dust attenuation curve in the UV-NIR range, for nearby galaxies, have been also measured as a function of either physical and structural parameter of the galaxies as the specific SFR (sSFR) and axis ratio or galaxy inclination and effective optical depth \citep[see e.g.,][with more active galaxies having shallower dust curves for increasing optical depths]{Wild2011b, Chevallard2013, KriekConroy2013}. All these works go in the direction of confirming the non-universality of the dust attenuation law.

In this paper, we want to investigate the UV-to-NIR shape of the dust attenuation curve of IR selected and luminous sources at high redshift.
We focus our attention on the sample of $z\sim2$ (U)LIRGs studied in \citet{LoFaro2013} for which a wealth of photometric and spectroscopic data and full radiative transfer analyses are available.
In order to characterize the shape of the dust attenuation curve we implement several prescriptions for dust attenuation, both well-known and new, within the context of state-of-the-art physically motivated spectral synthesis techniques and compare the results with those derived from theoretical modelling. We take advantage here of the high flexibility offered by the CIGALE code \citep[][]{Noll2009, Ciesla2014, Buat2014, Buat2015} which allows the user to implement his own prescriptions very easily. This work is performed within the context of the $Herschel$ Extragalactic Legacy project (European Union FP7-SPACE HELP\footnote{http://herschel.sussex.ac.uk/}) whose main goal is to assemble a new rich set of data and value-added parameters, (e.g. stellar and dust mass, SFR, AGN contribution etc.), characterizing the physical properties of millions of galaxies in the distant Universe covering 1000 sq. deg. or 1/40th of the entire sky. It is conceived to bring together observations from many  astronomical  observatories  to  provide  an  integrated dataset  covering  a  wide  range  of  wavelength  from  the  radio to the X-ray.

The sample used in this work is part of the ``pilot'' sample used within HELP to test and compare the different methods available to retrieve the physical properties of galaxies.

The paper is organized as follows. In Section 2, we present the reference sample of z$\sim$2 (Ultra) Luminous IR Galaxies ((U)LIRGs) used for this analysis. In Section 3 we present the methods used in this work to investigate the shape of the dust attenuation curve of high-$z$ IR bright sources. A detailed description of the physically-motivated SED-fitting approach is given with particular attention to the assumed parametrization for the dust attenuation curve. In Section 4 we first present the general results from SED-fitting related to the statistical analysis and parameter determination. We then present and discuss the attenuation curves estimated with CIGALE including the comparison with radiation transfer (RT) model results. In Section 5 we check the validity of our results for IR bright sources selected at lower redshifts. In Section 6 we explore the effects of the assumed parametrization for the dust attenuation curve on the derived SFR and M$_{\star}$ of galaxies. Our summary and conclusions are then presented in Section 7.

\section{Sample Selection}

As we are interested here in investigating the detailed shape of the attenuation curve of high-$z$ (U)LIRGs, we focus on a small but well-defined set of $Herschel$ sources selected to meet the following criteria: (1) full multiwavelength coverage from UV to sub-mm (including deep $Herschel$ observations from both PACS \citep{Poglitsch2010} and SPIRE \citep{Griffin2010} instruments  for the highest redshift bins);  (2) available spectroscopic redshifts; (3) galaxies powered by star formation (no AGN); and (4) available well-constrained radiative-transfer based solutions for these objects.

Our reference sample consists of 20 z$\sim$2 (U)LIRGs selected from the original sample described in \citet{LoFaro2013}. It includes the faintest 24 $\mu$m ($S_{24}\sim$  0.15-0.45 mJy) sources observed, in GOODS-S, with the Spitzer Infrared Spectrograph (IRS) by \citet{Fadda2010} with redshift in the interval 1.75 - 2.4. This specific redshift range allows us to sample the major contributors to the cosmic infrared background at the most active epochs. The selected galaxies are massive systems forming stars actively although not in a starburst regime. We refer to  \citet{LoFaro2013} for a   more detailed description of the sources. For the purpose of this work, the main characteristic of the sample is that it is crudely luminosity selected and that no other selection criterion has been applied.

All these galaxies benefit for high quality spectroscopic redshift measured from their observed IRS spectra by \citet{Fadda2010}.
In addition to the ultra-deep mid-IR spectra, these galaxies also benefit for a very rich suite of photometric data spanning a wide range in wavelengths from far-UV to Radio. These include far-UV to mid-IR broad band fluxes from the MUSIC catalogue by \citet{Santini2009}, full deep $Herschel$ imaging data from both SPIRE and PACS, covering the wavelength range between 70 and 500 $\mu$m and taken, respectively, from the $Herschel$ Multitiered Extragalactic Survey \citep[HerMES;][]{Oliver2012} and the PACS Evolutionary Probe \citep[PEP;][]{Lutz2011} programmes and radio VLA data at 1.4 Ghz. Typical noise levels in $Herschel$ bands are $\sim 1$ mJy for PACS $70-160$ $\mu$m and $\sim 6$ mJy for SPIRE $250-500$~$\mu$m, including confusion.The collected photometry, from UV to sub-mm, spans a maximum of 19 wavebands for all the sources. All the objects classified as AGN-dominated, on the basis of several indicators such as broad and high ionization lines in optical spectra, lack of a 1.6 $\mu$m stellar bump in the SED, X-ray bright sources, low mid-IR 6.2 $\mu$m equivalent width and optical morphology, have been carefully removed from the sample by   \citet{LoFaro2013}.

\section{SED modelling with energy balance: general features and assumed configurations}

The analysis performed in this work makes use of state-of-the-art spectral synthesis techniques which allow us to extract and interpret all the information contained in the observed integrated SEDs of galaxies in terms of their main physical properties and galaxy evolution in general \citep{Walcher2010}. Particular attention is payed to a proper treatment of dust effects. We make use, here, of SED-fitting techniques based on energy balance \citep[e.g.,][]{daCunha2008,Noll2009} to study the shape of the dust attenuation curve of our high-$z$ (U)LIRGs and then we compare their results to full radiative transfer computations.

The analysis is performed taking advantage of the high flexibility offered by
the code CIGALE \footnote {http://cigale.lam.fr}
 \citep[Code Investigating GALaxy Emission, developed in its first version with energy balance by][]{Noll2009}, to fit the spectral energy distributions of our high-$z$ (U)LIRGs.
The code allows one  to derive the main physical properties of galaxies from the observed shape of the UV-to-sub-mm SED by modelling their dust-reprocessed IR emission consistently with the UV-optical stellar SED making use of physically motivated IR templates. Here we are using an improved python version of the code (Boquien et al.~2017 in prep.). The modular structure of this new version of the code makes it very versatile for the use in different contexts and allows for higher flexibility in the SED modelling with the possibility for the user to implement his own prescriptions very easily. A description of this latest version of the code is given in \citet{Buat2015} and \citet{Ciesla2015}. Here we provide just a brief description of the code aimed at highlighting those features which have been particularly important to this analysis.

The CIGALE SED-fitting code is designed to provide, based on a Bayesian analysis, estimates of the mean values and uncertainties of the main physical parameters of the fitted galaxies
as well as the details on the assumed dust attenuation, stellar population(s) and AGN fraction (when required), evaluated from the dispersion of the probability distribution function (PDF) computed for each parameter. The reliability of the estimated parameters is also checked by creating mock galaxy catalogues where flux densities are built from the best model obtained for each source (plus an instrumental noise added to each flux) and analyzed using exactly the same method and modules used for the observed data.
A description of this method will be given in section \ref{mock-analysis}.

In the following subsections we briefly review the main ingredients of the code, including our own assumed configurations for each of its modules, with particular emphasis on the dust attenuation recipes considered in this work.

\subsection{Stellar component}

To build a galaxy SED we first need to define the properties of the underlying stellar population in terms of assumed stellar population library, Initial Mass function (IMF) and star formation history (SFH).

We adopt here the  stellar population models of \citet{BC03}
and a \citet{Salpeter1955} IMF (which is the same IMF adopted in the RT model used here for comparison).
Delay-$\tau$ SFHs, with varying e-folding time, are assumed to model the SFHs of our z$\sim$2 star forming (SF) galaxies. These are consistent with our galaxies being selected from the main sequence by \citet{Fadda2010} and are nowadays the most currently used to explain the SEDs of high-redshift star forming galaxies \citep[e.g.,][]{Lee2010, Maraston2010, Wuyts2011, Pforr2012}. Their functional form closely resembles that one derived from galaxy chemical evolution models and the RT model used here for comparison (see Section \ref{rt_modelling}).

The adopted parameters used, for the stellar component, in our fitting procedure are presented in Table \ref{tab-1}.

\subsection{Dust component: emission}
\label{ir_emission}

The energy absorbed by dust in the UV-optical is  re-emitted at longer wavelengths in the mid-IR to sub-mm range.

In CIGALE the IR emission of star-forming galaxies can be estimated by fitting their mid-IR- to sub-mm SEDs with either multi-parameter IR templates \citep[e.g.,][]{DraineLi2007} or with one-parameter templates \citep[e.g.,][]{Dale02} particularly suited when a good sampling of the  IR SED is not available.

In this work, the IR emission of our galaxies is modeled by fitting the dust models of \citet{DraineLi2007}, whose validity in reproducing the far-IR properties of $Herschel$ detected high-$z$ main sequence galaxies has been confirmed by the study of \citet{Magdis2012}. In these models the interstellar dust is described as a mixture of carbonaceous and amorphous silicate grains, whose size distributions are chosen to reproduce the observed extinction law in the Milky Way (MW), the Large Magellanic Cloud (LMC), and the Small Magellanic Cloud (SMC) bar region. The properties of these grains are parametrized by the polycyclic aromatic hydrocarbon (PAH) index, $q_{\rm PAH}$, defined as the fraction of the dust mass in the form of PAH grains.  \citet{DraineLi2007} assume that the majority of the dust is located in the diffuse interstellar medium (ISM), which is heated by a radiation field with a constant intensity $U_{\rm min}$. A smaller fraction $\gamma$ of the dust, representing the dust enshrouded in photodissociation regions (PDRs), is exposed to starlight with intensities ranging from $U_{\rm min}$ to $U_{\rm max}$ characterized by power-law distribution of the form, $dM/dU \propto U^{-\alpha}$.
{\citet{Draine2007} showed that the overall fit  of the SEDs of local  galaxies is insensitive to the adopted dust model (MW, LMC, and SMC) and the precise values of $\alpha$ and $U_{\rm max}$. Fixed values of $\alpha$ = 2 and $U_{\rm max}$ = $10^{6}$ can successfully reproduce the SEDs of the SINGS galaxy sample. The same values have been used by \citet{Magdis2012} to reproduce the IR properties of a sample of high-$z$ main sequence galaxies.
Following \citet{Magdis2012} we have considered, in this work, MW dust models with $q_{\rm PAH}$ ranging from 1.12\% to 3.19\%, which are the typical values for high-$z$ normal star forming galaxies,  and U$_{\rm min}$ varying in the interval 5 - 25.
The value of the fraction $\gamma$ of dust enclosed in the PDRs has been fixed to the mean value obtained by \citet{Magdis2012} for the observed average SEDs of main-sequence galaxies at redshift 1 and 2 derived by stacking analysis (i.e. $\gamma$=0.02).

The final adopted configuration for the \citet{DraineLi2007} models is shown in Table~\ref{tab-1}.

\subsection{Prescriptions for dust attenuation}
\label{attenuation_curves_recipes}

When modelling the SEDs of galaxies we have seen that the effect of dust  depends on the combination of both the optical and physical properties of dust and the relative geometrical distribution of stars and dust. In CIGALE all these complex effects can be  accounted for by assuming flexible attenuation curves with varying shape according to the specific case under analysis. In order to account for the large variety of objects and possible geometrical configurations we have implemented in CIGALE a new recipe for  dust attenuation, called hereafter DouBle-Power law model (\textit{DBPL-free}), where the UV-to-NIR shape of the curve is treated as a free parameter, and we have compared this recipe with several well-known standard prescriptions for  dust attenuation. All the different recipes for the dust attenuation considered in this work are here below described, starting from the standard ones already included in CIGALE and ending with the newly implemented DBPL-free model.

\subsubsection*{3.3.1 Calzetti-like recipes}
\label{calzetti}

The \citet{Calzetti2000} attenuation law, initially calibrated on a sample of 39 nearby UV-bright starburst galaxies \citep[][]{Calzetti1994}, is nowadays the most currently used also at high redshift.
It is characterized by a grayer slope than the MW and LMC extinction laws accounting for the effects of geometry and mixing compared to simple extinction.

The attenuation is parametrized as:
\begin{equation}
f_{\rm int}(\lambda) = f_{\rm obs}(\lambda)10^{0.4 k^{\rm e}(\lambda) E(B-V)_{\rm star}}
\end{equation}
where $k^{\rm e}(\lambda)$ is the effective attenuation curve to be applied to the observed stellar continuum SED $f_{obs}(\lambda)$ of a starburst galaxy to recover the intrinsic SED $f_{int}(\lambda)$ with
\begin{equation}\begin{split}
k^{\rm e}(\lambda) = 2.659 (- 1.857 + 1.040/\lambda) + R_{\rm V} \\
 (0.63 \mu {\rm m} \leq \lambda \leq 2.20 \mu {\rm m} ) \\
                        = 2.659 (-2.156 + 1.509/\lambda - 0.198/\lambda^{2} + 0.011/\lambda^{3}) + R_{V}\\
                        (0.12 \mu {\rm m}  \leq \lambda < 0.63 \mu {\rm m} )
\end{split}\end{equation}
and $R_{\rm V} = A_{\rm V}/E(B-V)$ = 4.05.\\

CIGALE includes the possibility to modify the \citet{Calzetti2000} attenuation curve by varying the steepness of the law in the UV-optical range and by adding a bump
centered at 2175 \AA~ using a Lorentzian-Drude profile according to the following formalism:
\begin{equation}
A(\lambda) = \frac{A_{\rm V}}{4.05} (k^{\prime}(\lambda) + D_{\lambda_{0},\gamma,E_{\rm b}}(\lambda))(\frac{\lambda}{\lambda_{\rm V}})^{\delta}
\label{dust1}
\end{equation}
where $\lambda_{\rm V} = 5500$ \AA,~$k^{\prime}$  comes from \citet{Calzetti2000} and  $D_{\lambda_{0},\gamma, E_{\rm b}}$ is the Lorentzian-like Drude profile \citep[][]{Fitzpatrick1990, Noll2009}.

In complement to the exact Calzetti recipe we implement in CIGALE the specific recipe described in \citet{Buat2011, Buat2012}. The authors, by analyzing a well defined UV selected and $Herschel$ detected sample of strong UV emitting galaxies at z$>$1, found evidence for a steeper UV slope than Calzetti and for a 2175 $\AA$ bump.
Following \citet{Buat2011} we fix the slope of the power-law modifying the Calzetti law, $\delta$, to -0.13. The parameters of the bump, namely the peak amplitude above the continuum, FWHM and central wavelength of the bump, ($E_{\rm b}$, $\gamma$  and $\lambda_{0}$), are fixed to 1.26, 35.6 nm and 2175 \AA~, respectively.   Note that we do not study the presence or not of a bump in this work and that we introduce the \citet{Buat2011}  recipe to check the impact of a law steeper than the Calzetti one but based on the same parametrization.

In CIGALE the same attenuation law is assumed for both young \footnote{We fix the separation age between young and old stellar populations to 10 Myrs, as in \citet{CharlotFall00}} and old stars but a reduction factor of the visual attenuation (expressed in magnitude),  $f_{\rm att}=A_{V}^{\rm ISM}/A_{V}^{\rm BC}$, is applied to the old stellar population in order to account for the \textit{age-dependent} attenuation of stars within a galaxy \citep[][]{Calzetti1994, Calzetti2000, CharlotFall00, Panuzzo2007, Buat2011, Buat2012}.

Current studies on the dust properties of high redshift galaxies have led to contrasting results concerning the $f_{\rm att}$ parameter. Some authors have confirmed the validity of local Universe \citet{Calzetti2000} prescription for  a  higher  color excess to be applied to the gas recombination lines than to the stellar continuum, even at high-$z$ \citep[e.g.,][]{ForsterSchreiber2009, Whitaker2014, Yoshikawa2010}. On the other hand a comparable attenuation for the young (emission lines) and old (stellar continuum) component have been estimated in UV selected z$\sim$2 galaxies by other authors \citep[e.g.,][]{Erb06, Reddy2010}. Similar results have been recently found by \citet{Puglisi2016} for a sample of IR selected and luminous sources at z$\sim$1 by comparing the dust attenuation of continuum stellar emission and nebular emission from $Herschel$ and H$\alpha$ fluxes. A decrease of the amount of extra attenuation towards star forming regions with respect to diffuse dust as a function of increasing sSFR has been observed by \citet{Wild2011b} and successively by \citet{Price2014} for both local and higher redshift galaxies. This has been interpreted by the authors as a direct evidence of the two component dust model \citep[e.g.,][]{Calzetti1994, CharlotFall00, Granato2000}. In fact, in galaxies with higher specific SFRs, the continuum light is mostly contributed by young massive stars in the birth clouds, so both the continuum and emission lines may be attenuated by both dust components, thus bringing to higher $f_{\rm att}$ on average.

To account for the different observational evidences in our analysis this attenuation factor is treated as a free parameter and five different values between 0.3 and 1.0 are considered. These are listed in Table~\ref{tab-1}.

\subsubsection*{3.3.2 Double power law attenuation recipes}

\label{DBPLfree}

A complementary method to define an attenuation law is provided by the simple model of \citeauthor{CharlotFall00}~(2000, CF00 hereafter). In their model for dust attenuation, young stars are embedded in dusty BCs, which dissolve over a finite timescale of  $\sim$ 10$^{7}$ yr, and, in addition, all stars experience dust attenuation due to the diffuse ISM. They define an effective absorption optical depth  $\tau_{\lambda}^{'}$\footnote{CF00  introduce a  transmission function at a given wavelength by $T_{\lambda} = \exp(-\tau_{\lambda})$,  $\tau_{\lambda}$ being the effective absorption optical depth.} whose time dependence is formalized as:
\begin{eqnarray}
 \tau_{\lambda}(t') =
  \begin{cases}
 \tau_{\lambda}^{\,\mathrm{BC}}+\tau_{\lambda}^{\,\mathrm{ISM}} & \text{for }  t^{\prime} \leq t_{0} \\
 \tau_{\lambda}^{\,\mathrm{ISM}}     & \text{for }  t^{\prime} > t_{0}.
  \end{cases}
\end{eqnarray}
with $\tau_{\lambda}^{\rm BC} $ being the effective absorption optical depth of the dust in the birth clouds and $\tau_{\lambda}^{\rm ISM}$ that one in the diffuse ISM.
The shape of the attenuation curve is assumed to be a power law function of wavelength, with a normalization that depends on the age of the stellar population:
\begin{flalign}
\label{taubc_cf00}
& \tau^{\rm BC}_{\lambda} = (1 - \mu)\tau_{\rm V} (\lambda/5500 \AA)^{\delta_{\rm BC}}\\
& \tau^{\rm ISM}_{\lambda} = \mu \tau_{\rm V} (\lambda/5500 \AA)^{\delta_{\rm ISM}}
\label{tauism_cf00}
\end{flalign}
with $\tau_{\rm V}$ representing the total V-band attenuation experienced by young stars within the birth molecular clouds (due to the BCs themselves and the diffuse ISM) and $\mu$ being the fraction of the total effective optical depth contributed by the diffuse ISM.

Starting from the formalism of CF00, we have developed in CIGALE a new module called ``\textit{dustatt\_2powerlaws}'' which  allows  to combine different slopes for the attenuation curve relative to the ISM and BC component. In this way we are able to properly account for an age-dependent attenuation where not only the total amount of dust attenuation changes as a function of the stellar age but also the way the stellar light is extinguished,   as originally introduced by CF00. To be consistent with the formalism adopted in CIGALE for the Calzetti-like recipes, we use the parameter $f_{\rm att}$ ($A_{\rm V}^{\rm ISM}/A_{\rm V}^{\rm BC}$) instead of $\mu$ ($\tau_{\rm V}^{\rm ISM}$/($\tau_{\rm V}^{\rm BC} + \tau_{\rm V}^{\rm ISM}$)). The values of $f_{\rm att}$ reported in Table~\ref{tab-1} are equivalent to values of the $\mu$ parameter in the interval 0.2-0.5 (being $\mu = f_{\rm att}/(1+f_{\rm att})$). We also use the total V-band attenuation experienced by young stars within the molecular clouds, $A_{\rm V}^{\rm BC}=1.086 \tau_{\rm V}$ in place of $\tau_{\rm V}$. The latter comes from the fact that, in CF00 formalism, $L_{\lambda}/L_{\lambda}^{0} \equiv \exp(-\tau_{\lambda})$ and $A_{\lambda} \equiv -2.5 \log (L_{\lambda}/L_{\lambda}^{0})$.

We implement through our \textit{dustatt\_2powerlaws} module three different configurations here below itemized:
\begin{itemize}
\item \textit{Standard CF00}: here the slope of the attenuation curve is assumed, for simplicity, to be the same for the BC and ISM component and it is fixed to the canonical value of $\delta_{\rm ISM} = \delta_{\rm BC}$ = -0.7.
This value allows the authors to successfully reproduce the observed mean relation between  IRX and UV spectral slope of nearby starburst galaxies \citep[see e.g.,][]{Calzetti1994, Calzetti2000, CharlotFall00, Johnson2007}.
\\
\item \textit{MAGPHYS-like}: we consider here the recipe used in the semi-empirical SED-fitting code, MAGPHYS (Multi-wavelength Analysis of Galaxy Physical Properties), of \citet{daCunha2008}, hereafter referred as ``\textit{MAGPHYS-like}'' configuration.

In this case the same slope as in CF00 is adopted for the ISM component ($\delta_{\rm ISM}=-0.7$) while a steeper slope, $\delta_{\rm BC}=-1.3$, is assumed for the dense birth clouds following the original recipe first proposed by Charlot \& Fall.
\\
\item \textit{DBPL-free}: this represents our reference recipe in this work. Its novelty is the possibility to treat both the slopes of the ISM and BC dust attenuation curves as free parameters. Our configuration includes both steeper and grayer slopes (compared to CF00 classic recipe) for $\delta_{\rm BC}$ and grayer values for $\delta_{\rm ISM}$. More specifically the BC and ISM slopes can assumed the values (-1.3, -0.7, -0.5) and (-0.7, -0.5, -0.3, -0.1), respectively.

For what concerns the BC component, despite the wide use of -1.3 as reference slope for the extinction curve of the single BCs \citep[e.g.,][]{CharlotFall00, daCunha2008, Wild2011a, Wild2011b, Chevallard2013}, evidences in favor of a grayer slope are also shown. For example, by studying the extinction and reddening of LMC HII regions, \citet{Caplan1986} found clear evidence for a total-to-selective dust extinction ratio $R_{\lambda_{\rm H_{\beta}}} = 5.1$ corresponding to a  slope for the extinction curve of LMC HII regions of $\sim -0.74$ which is very close to the value adopted by CF00. A clumpy dust geometry was in this case invoked to explain the observed slope. \citet{Liu2013} have also shown that at the scale of local HII regions (M83), grayer extinction curves coupled with more complex geometries are expected while when averaged over large scales ($> 100-200  $ pc), the extinction becomes consistent with a ``dust screen'' geometry. Hidden emission from young stars embedded in dense molecular clouds (MCs) can also be responsible for grayer attenuation curves, on average, in particular when dealing with Ultra luminous IR galaxies characterized by SFR higher than $\sim$ 200 M$_{\odot}$/yr \citep[][]{Flores2004}.

The shallowness of the attenuation curve at large optical depths for the ISM component is a clear signature of mixed distributions of stars and dust \citep[e.g.,][]{CharlotFall00, Gordon2000, Calzetti2001, Pierini2004, Chevallard2013}. The classical value of $\delta_{\rm ISM} = -0.7$ adopted in CF00 model was specifically constrained to reproduce \citet{Calzetti2000} recipe.    We have already seen in previous sections that deviations from the Calzetti relation are expected for both nearby and high-$z$ IR luminous galaxies which lead us to explore a wider range of slopes including grayer values than -0.7.
\end{itemize}

\begin{figure}
\centerline{
\includegraphics[width=8.6cm]{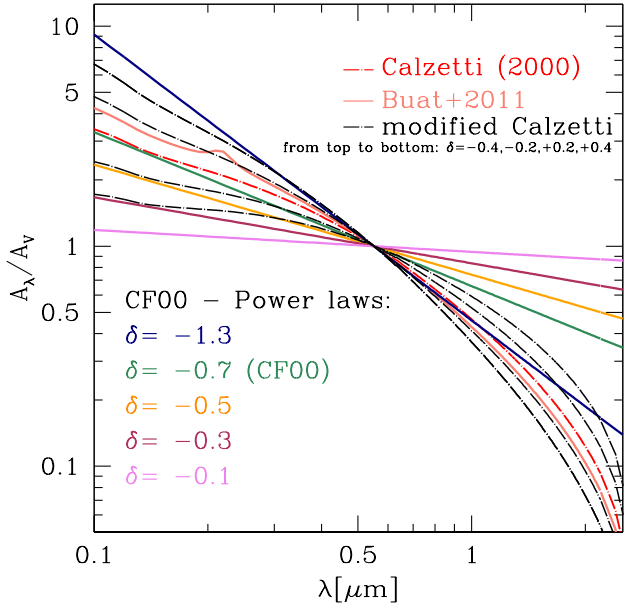}}
\caption{All the assumed wavelength dependent shapes for the dust attenuation curves considered in our analysis are shown in this Figure. As red long-dashed line the \citet{Calzetti2000} attenuation law, as salmon solid line the modified Calzetti curve based on the formalism defined in \citet{Buat2011}, then color-coded from blue to violet and from top to bottom the power law slopes assumed for the CF00 based recipe are also shown. Note that here we are showing the single power-law dependence assumed for the two, BC and ISM, components and not the combined curve obtained from their combination. The latter, in fact, depends on the physical properties of the galaxy and the relative proportion of old and young stars. Three different slopes have been assumed for the BC component (-1.3, -0.7, -0.5) and four for the ISM (-0.7, -0.5, -0.3, -0.1). The large set of different shapes for the attenuation curve considered here includes both steeper and grayer slope than the original Calzetti  recipe both at short and long (NIR) wavelengths. All the recipes included in this analysis and described by the formalism specified in section~\ref{attenuation_curves_recipes} are then compared with the modified-Calzetti recipes (black dashed lines) used in \citet{Salmon2015} to study the UV shape of the attenuation curve of CANDELS galaxies at redshift 1.5-3 and their dependence on the color excess E(B-V). The comparison shows that we cover the same range of dust attenuation slopes in the UV while the major differences are in the NIR, whose investigation is the main objective of this work.}
\label{attenuation_curves_recipes_fig}
\end{figure}
All the different recipes for dust attenuation considered here and discussed in previous sections are shown in Figure~\ref{attenuation_curves_recipes_fig}. To highlight the main differences between the two complementary formalisms presented here for the dust attenuation curve, namely the modified Calzetti and the double power-law one, we also plot in the figure the curves corresponding to the modified-Calzetti recipes (as black dashed lines) used in \citet{Salmon2015} to study the UV shape of the attenuation curve of CANDELS galaxies at redshift 1.5-3. Although both formalisms appear to be able to cover the same range of slopes in the UV they significantly differ at longer wavelengths where the double power-law recipes (color-coded lines) provide grayer slopes than modified-Calzetti ones.
.

Our DBPL-free recipe allows us to explore a large set of different shapes for the attenuation law including both steeper and grayer slopes than \citet{Calzetti2000} in the UV-optical and, more interestingly,  grayer slopes in the NIR too.

Table \ref{tab-1} finally summarizes all the assumed input parameter configurations, for both the stellar and dust component, used in the phenomenological modelling to investigate the physical properties of our IR-selected and luminous galaxies at high-$z$. The choice of a delay-$\tau$ SFH, characterized by only two free parameters, together with an assumed configuration for the dust emission where most of the parameters have been fixed and with dust attenuation recipes where maximum three parameters (as in the DBPL-free) are treated as free, bring to a total number of free parameters ranging between 5 and 7. The number of photometric datapoints is always larger than 18.

\begin{table*}
\scriptsize
\centering
\begin{tabular}{c c c}
\hline\hline
Parameter & Symbol & Range \\
\hline
\multicolumn{3}{c}{SFH (common to all the assumed dust configurations)}\\
\hline
age  & $age$ & 0.1, 0.5, 1., 1.2, 1.4, 1.8, 2., 2.2, 2.4, 2.8, 3., 3.2, 3.4, 3.8\,Gyr\\
$e$-folding timescale of delay-$\tau$ SFH & $\tau$ & 0.1, 0.5, 1., 1.5, 2., 2.5, 3., 7.\,Gyr\\
\hline
\multicolumn{3}{c}{Dust emission}\\
\hline
Mass fraction of PAH & $q_{\rm PAH}$ & 1.12, 2.50, 3.19\\
Minimum radiation field & U$_{\rm min}$ & 5., 10., 25.0\\
Powerlaw slope dU/dM $\propto$ U$^{\alpha}$ & $\alpha$ & 2.0\\
Dust fraction in PDRs & $\gamma$  & 0.02\\
\hline
\multicolumn{3}{c}{Dust attenuation}\\
\hline
\textit{Calzetti~(2000):} & &\\
Colour excess of stellar continuum light for young stars & $E(B-V)_{\rm young}$ & 0.12, 0.25, 0.37, 0.5, 0.62, 0.74, 0.86, 0.98, 1.10, 1.43, 1.6, 1.98\\
Reduction factor for the E(B-V) of the old stars compared to the young ones & $f_{\rm att}$ & 0.3, 0.5, 0.8, 1.0\\
\hline
\textit{Modified Calzetti (Buat et al.~2011):} & &\\
Colour excess of stellar continuum light for young stars & $E(B-V)_{\rm young}$ & 0.12, 0.25, 0.37, 0.5, 0.62, 0.74, 0.86, 0.98, 1.10, 1.43, 1.6, 1.98\\
Reduction factor for the E(B-V) of the old stars compared to the young ones & $f_{\rm att}$ & 0.3, 0.5, 0.8, 1.0\\
Central wavelength of the UV bump in nm & $\lambda_{0}$ & 217.5\\
Full width half maximum of the bump profile in nm & $\gamma$ & 35.6\\
Amplitude of the UV bump & $E_{\rm b}$ & 1.26\\
Slope of the power law modifying the attenuation curve & $\delta$  & -0.13\\
\hline
\textit{CF00 recipe (power law att. curve):} & &\\
V-band attenuation in the birth clouds (BCs) & $A_{\rm V}^{\rm BC}$ & 0.5, 1., 1.5, 2.0, 2.5, 3.0, 3.5, 4.0, 4.5, 5.8, 6.5, 8.0\\
Power law slope of dust attenuation in the BCs & $\delta_{\rm BC}$ & -0.7\\
$A_{\rm V}^{\rm ISM}/ A_{\rm V}^{\rm BC}$ & $f_{\rm att}$ & 0.3, 0.5, 0.8, 1.0\\
Power law slope of dust attenuation in the ISM & $\delta_{\rm ISM}$ & -0.7\\
\hline
\textit{Magphys recipe (DBPL):} & &\\
V-band attenuation in the birth clouds (BCs) & $A_{\rm V}^{\rm BC}$ & 0.5, 1., 1.5, 2.0, 2.5, 3.0, 3.5, 4.0, 4.5, 5.8, 6.5, 8.0\\
Power law slope of dust attenuation in the BCs & $\delta_{\rm BC}$ & -1.3\\
$A_{\rm V}^{\rm ISM}/ A_{\rm V}^{\rm BC}$ & $f_{\rm att}$ & 0.3, 0.5, 0.8, 1.0\\
Power law slope of dust attenuation in the ISM & $\delta_{\rm ISM}$ & -0.7\\
\hline
\textit{Double Power law free (DBPL-free):} & &\\
V-band attenuation in the birth clouds (BCs) & $A_{\rm V}^{\rm BC}$ & 0.5, 1., 1.5, 2.0, 2.5, 3.0, 3.5, 4.0, 4.5, 5.8, 6.5, 8.0\\
Power law slope of dust attenuation in the BCs & $\delta_{\rm BC}$ & -1.3, -0.7, -0.5\\
$A_{\rm V}^{\rm ISM}/ A_{\rm V}^{\rm BC}$ & $f_{\rm att}$ & 0.3, 0.5, 0.8, 1.0\\
Power law slope of dust attenuation in the ISM & $\delta_{\rm ISM}$ & -0.7, -0.5, -0.3, -0.1\\
\hline\hline
\end{tabular}
\caption{Input parameter configurations used in CIGALE to fit the z$\sim$2 (U)LIRGs}
\label{tab-1}
\end{table*}
\section{Fitting the spectral energy distributions: dust attenuation laws}
\label{Results}

We discuss in this section the results obtained by fitting the observed SEDs of our high-$z$ (U)LIRGs with CIGALE. The code is primarily run under our reference configuration (DBPL-free) and then for each of the parameter configurations listed in Table~\ref{tab-1} for comparison. The fit is performed by comparing the model SED integrated over a specific set of filters with the observed flux densities provided by the input catalogue. The best model is obtained by  minimizing the reduced  $\chi^{2}$ ( hereafter $\chi^{2}_{\nu}$) associated to each model and set of parameters. The input parameters as well as additional output parameters are estimated by building the probability distribution function (PDF) and
 by taking the mean and standard deviation of the PDF \citep[see][for a more detailed discussion about the methods used in CIGALE]{Noll2009}. The typical output of CIGALE includes, in addition to the parameters related to the dust attenuation curves, few other parameters such as SFR, M$_{\star}$, galaxy age and total dust attenuation.

Before discussing the attenuation curves presented in Section~\ref{discuss_att_curves}, we check the validity and accuracy of our estimates, also in terms of parameter degeneracy, through the mock analysis described below.

\subsection{Mock analysis and parameter determination}
\label{mock-analysis}

The mock analysis is performed by generating with CIGALE catalogues of artificial sources for which the physical parameters are known by definition \citep[details can be found in][]{Buat2012, Buat2014, Ciesla2015}.
To build the mock catalogue we use the best-fit model  of each of the studied objects previously obtained through our SED-fitting procedure. The flux densities of the mock SEDs are computed by randomly picking a flux value from the normal distribution generated using as mean value the best model flux and as standard deviation the photometric error.
\begin{figure*}
\centerline{
\includegraphics[angle=-90, width=6.5cm]{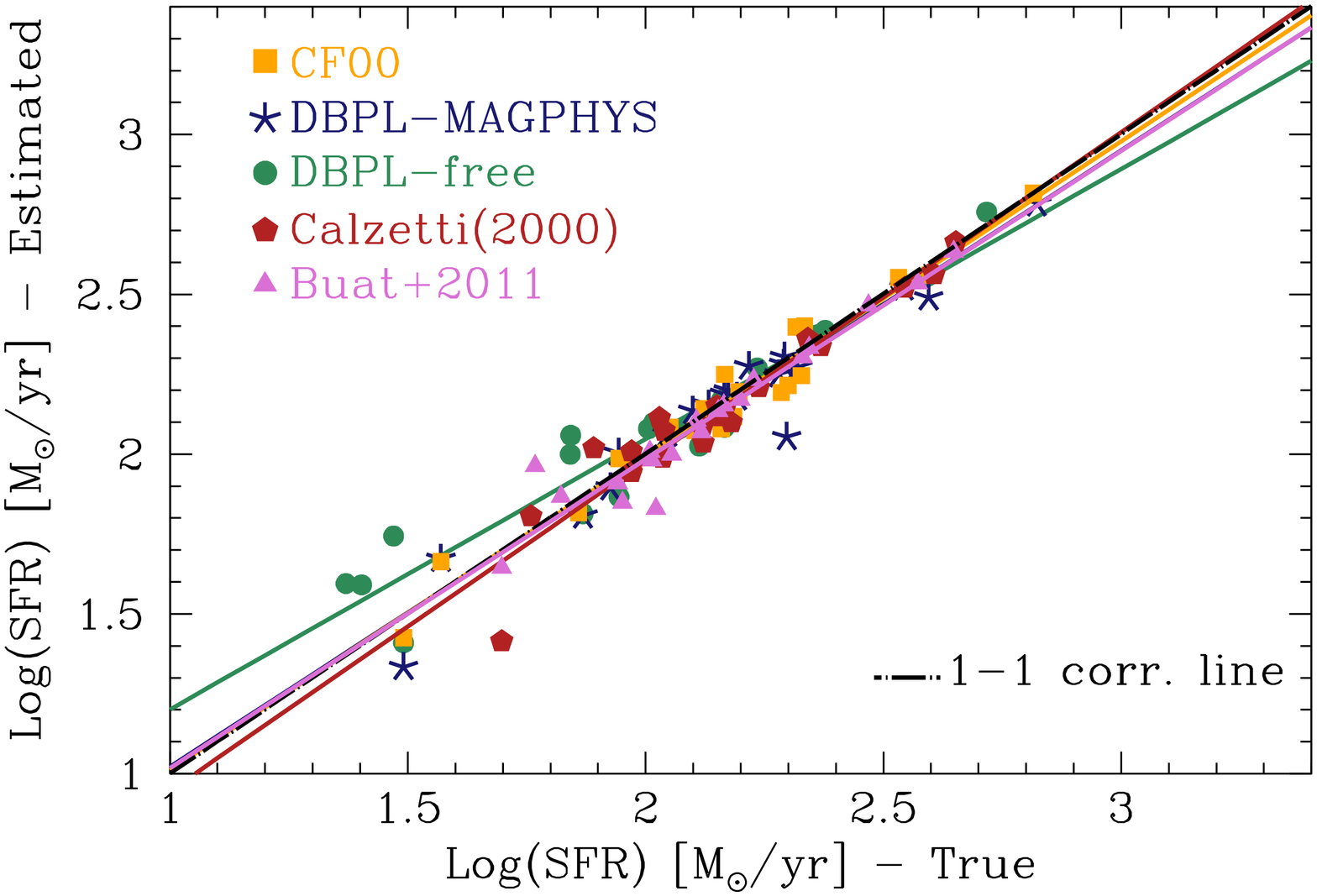}
\includegraphics[angle=-90, width=6.5cm]{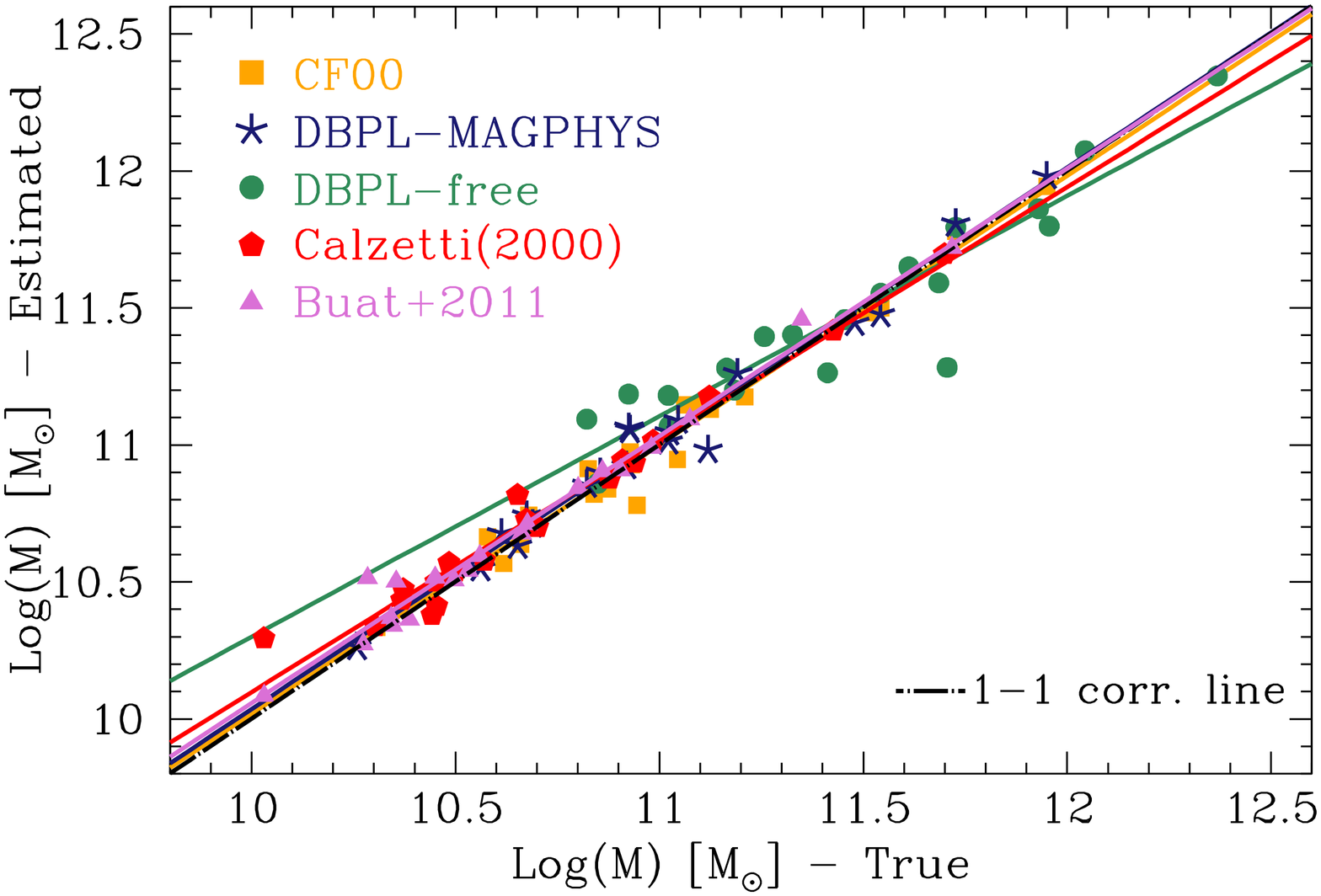}
\includegraphics[angle=-90, width=6.5cm]{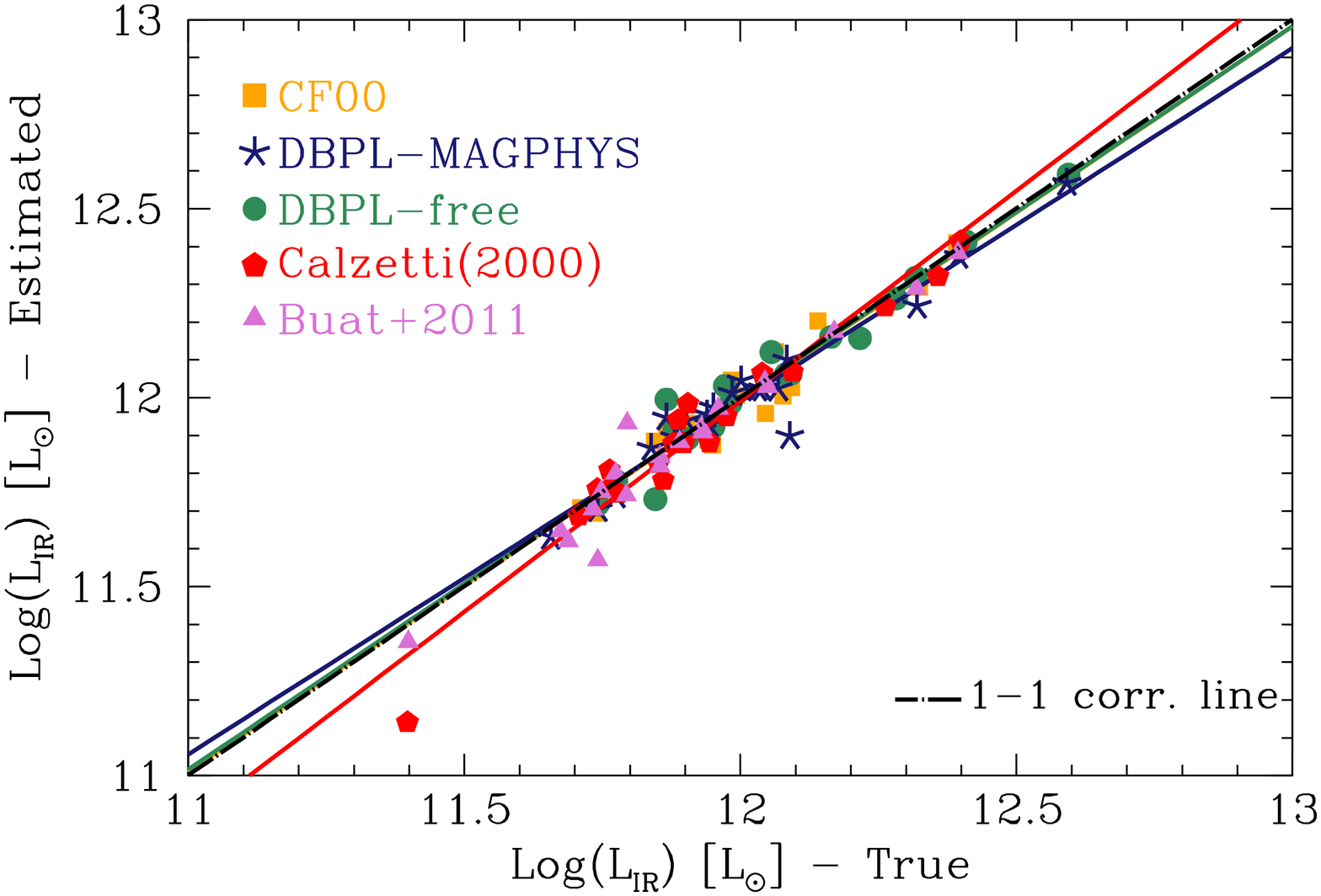}}
\centerline{
\includegraphics[angle=-90, width=6.5cm]{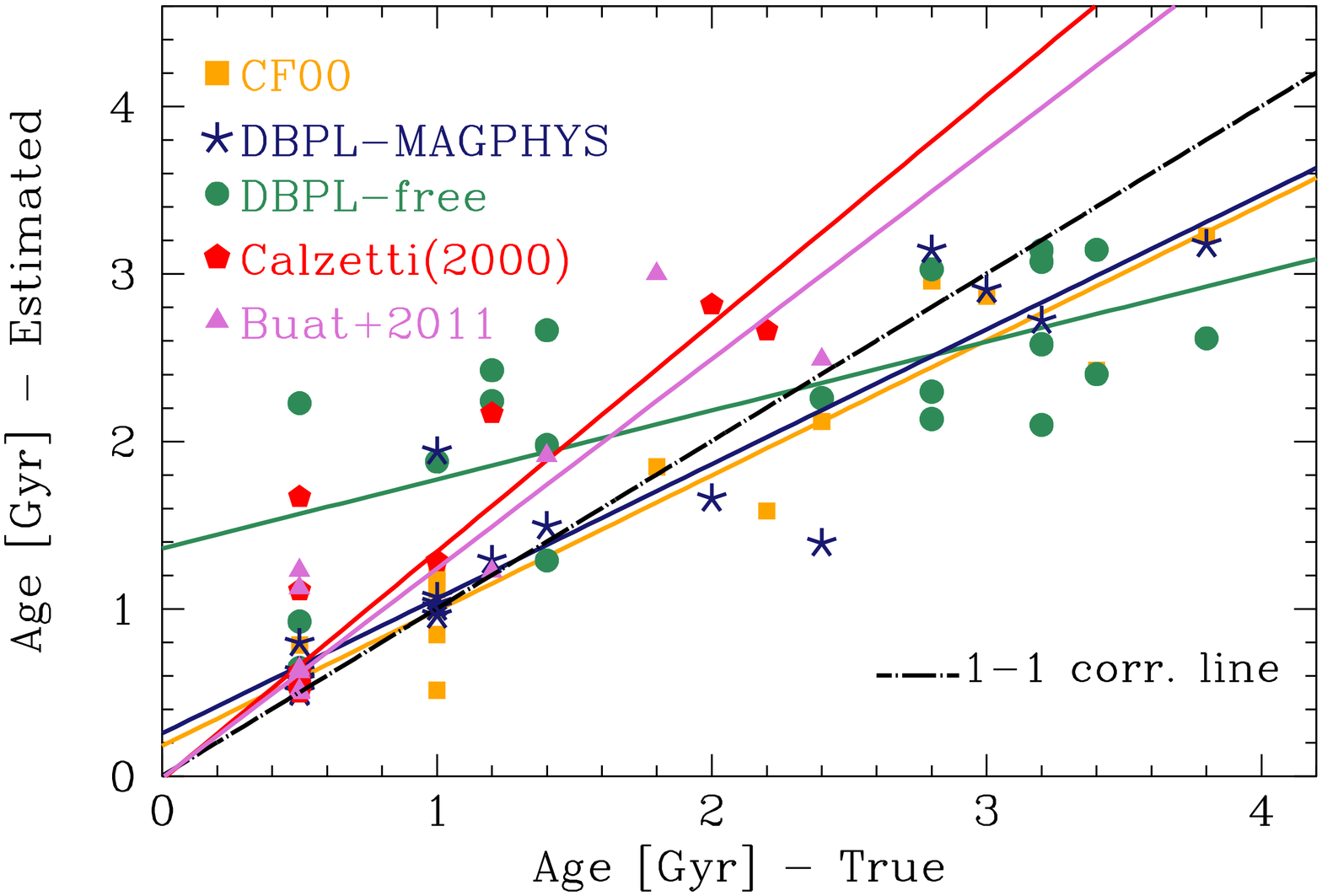}
\includegraphics[angle=-90, width=6.5cm]{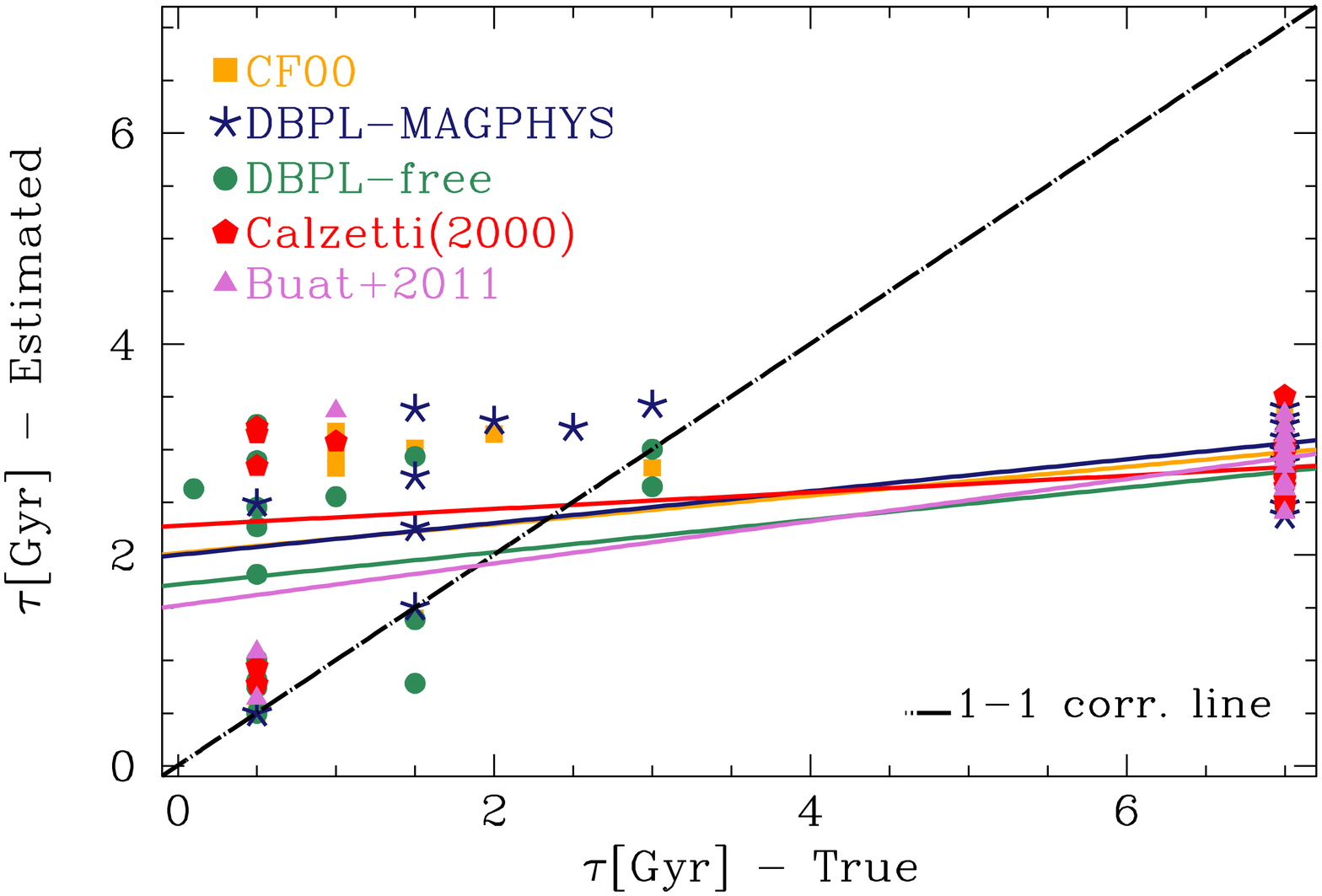}
\includegraphics[angle=-90, width=6.5cm]{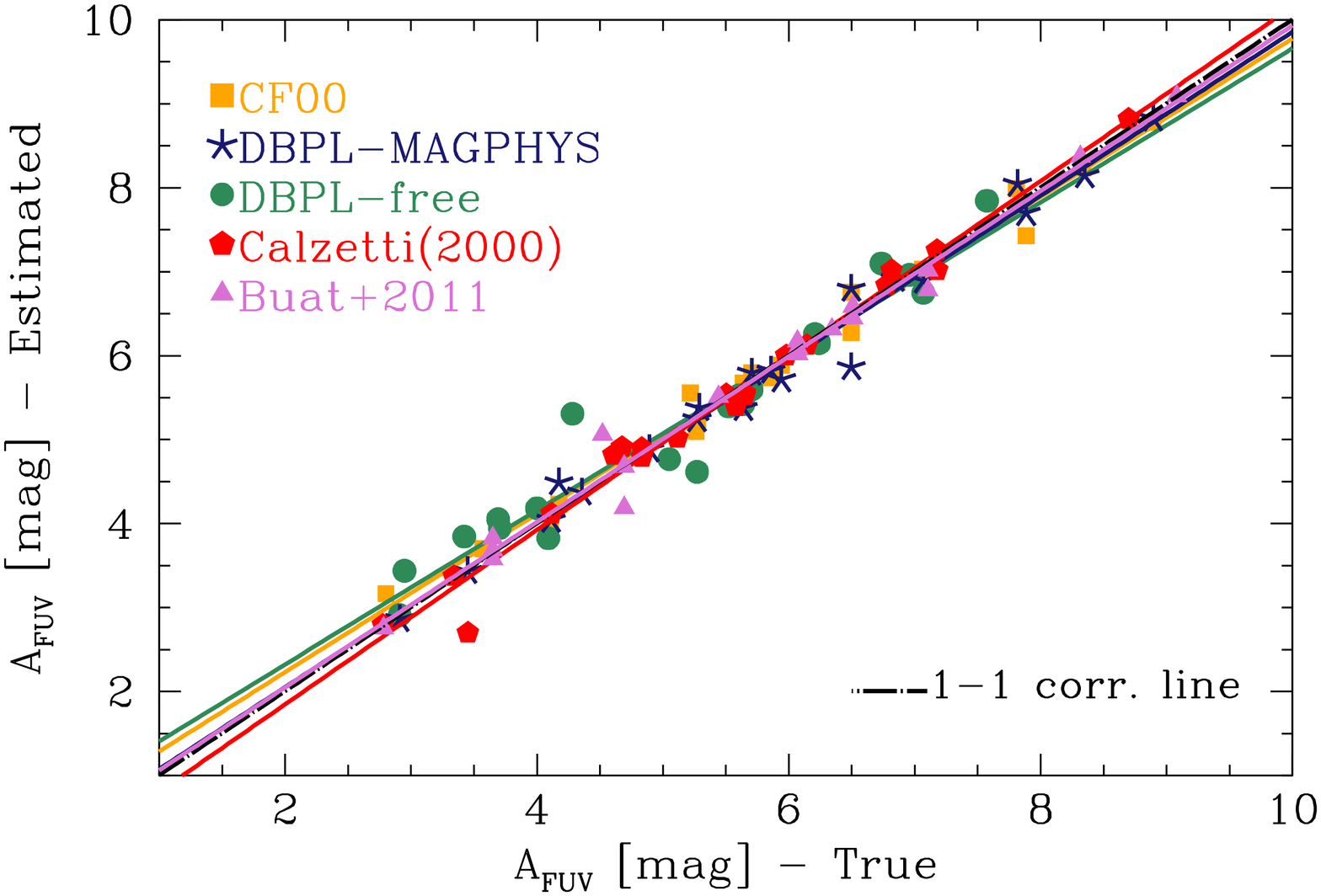}}
\centerline{
\includegraphics[angle=-90, width=6.5cm]{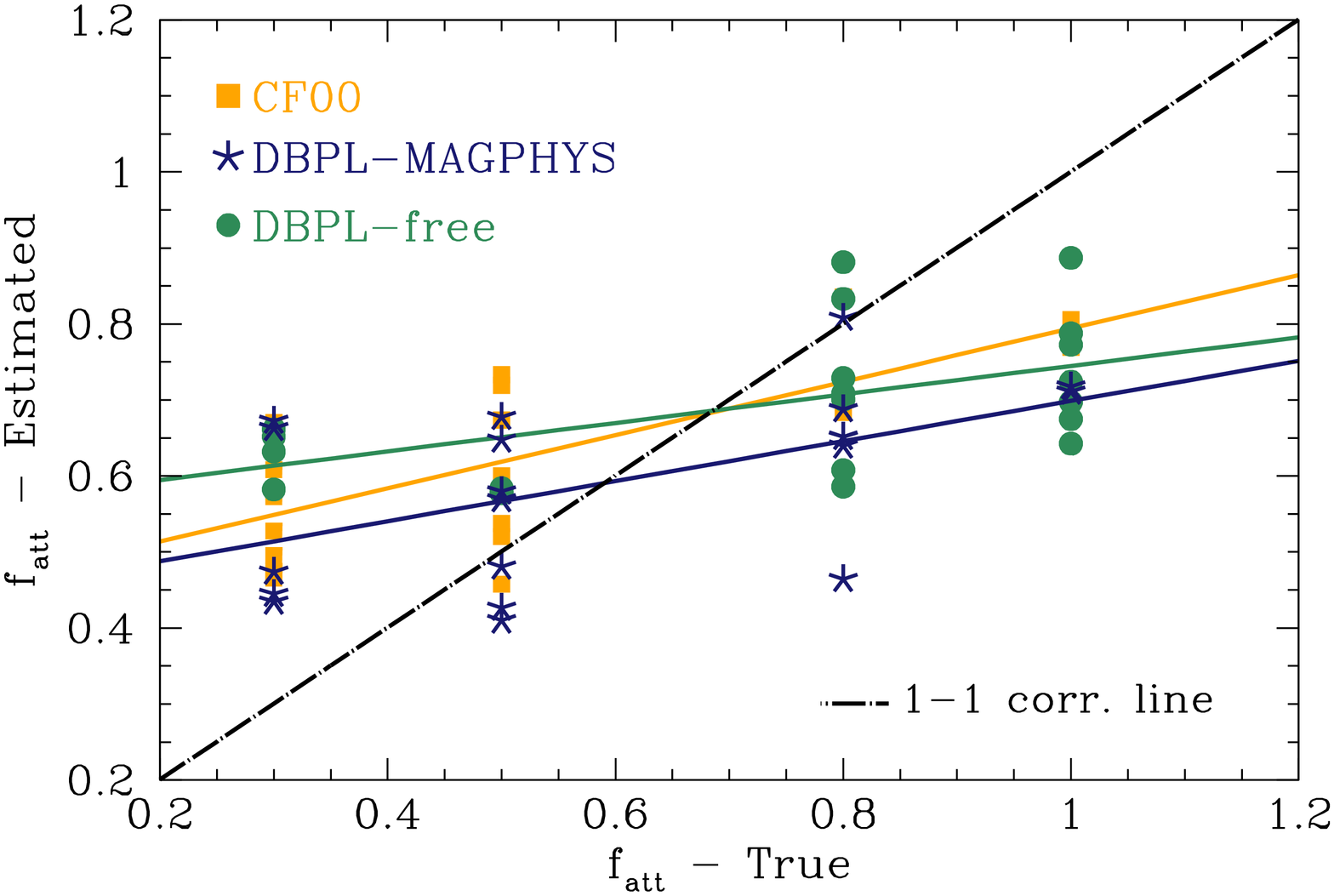}
\includegraphics[angle=-90, width=6.5cm]{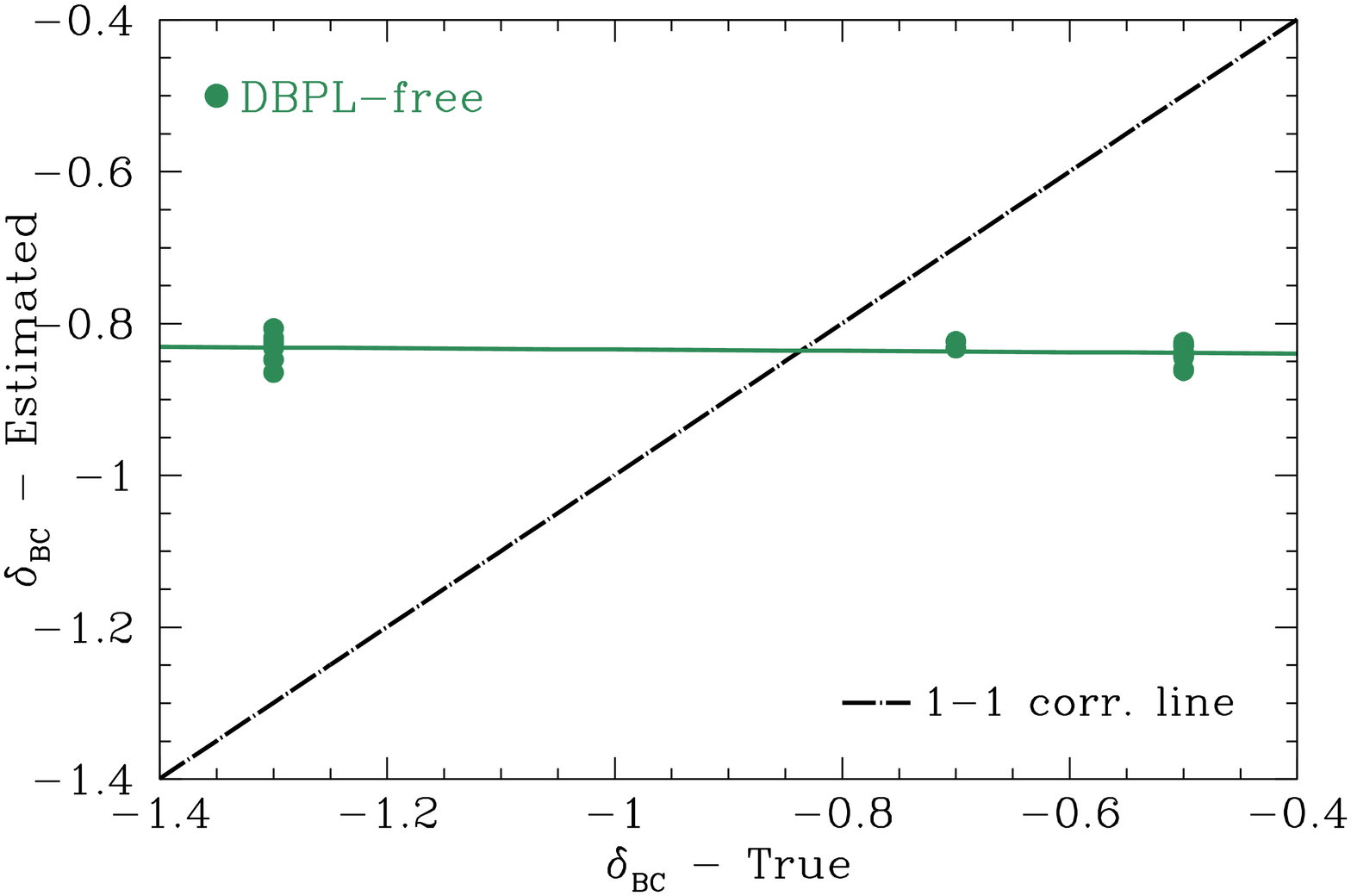}
\includegraphics[angle=-90, width=6.5cm]{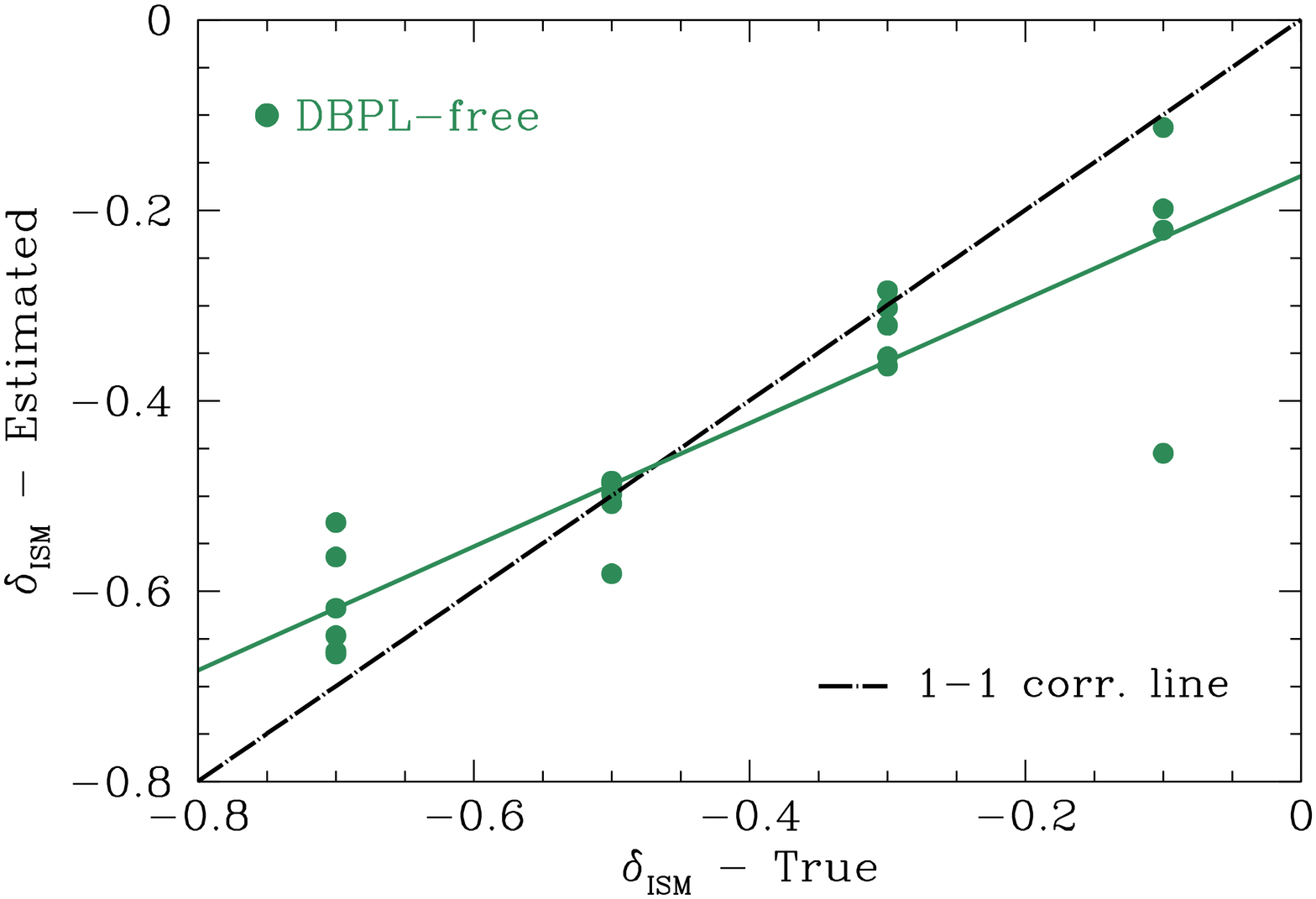}}
\caption{Comparison between the true value of the output parameters provided by our best-fit model (artificial SEDs), on the x-axis, and the same parameter estimated by the code on the y-axis. The 1-1 correlation line is shown as black long dashed-dotted line in each panel. From top left to bottom right the results for the star formation rate, stellar mass, total IR luminosity, age, e-folding timescale of delay-$\tau$ SFH, total FUV-Attenuation, BC-to-ISM attenuation factor $f_{\rm att}$ (definition in Section~\ref{calzetti}) and the slope of BC and ISM attenuation curves, are shown color coded according to the five different dust attenuation recipes listed in Tab.~\ref{tab-1}. The regression lines for each assumed configuration are also plotted as colored solid lines.}
\label{mock_analysis_fig}
\end{figure*}
\begin{figure}
\centering
\includegraphics[angle=-90, width=6.2cm]{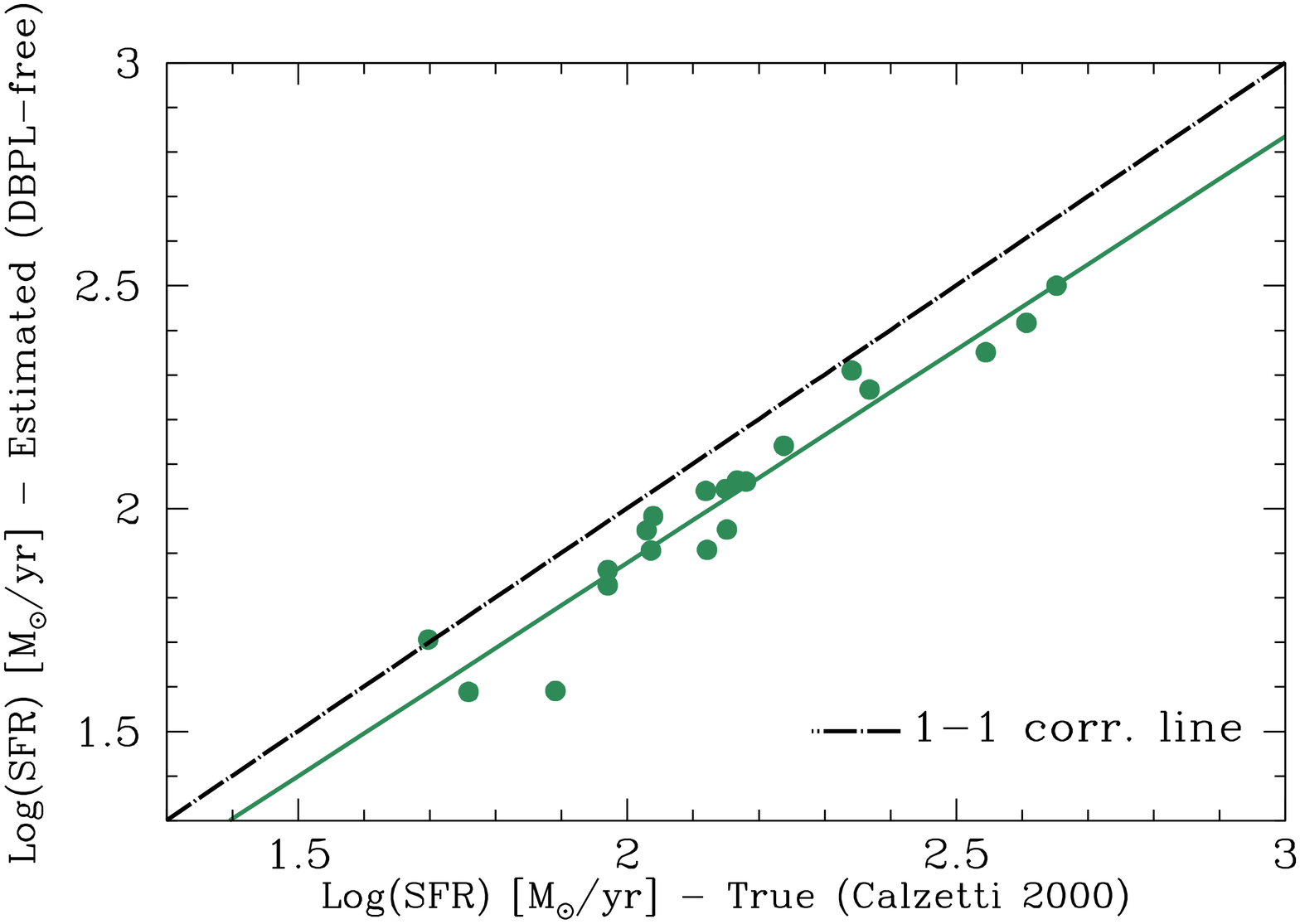}
\includegraphics[angle=-90, width=6.2cm]{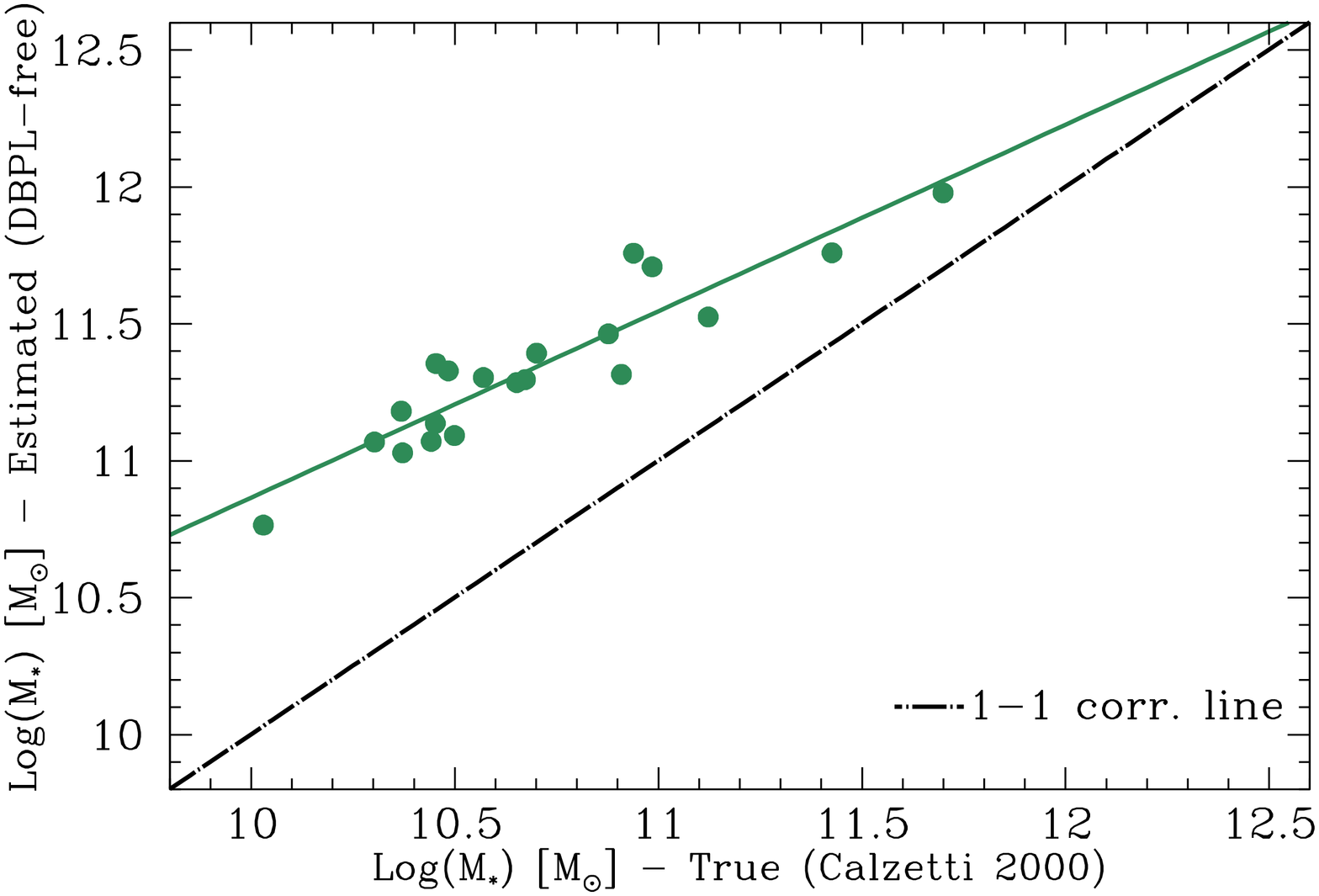}
\includegraphics[angle=-90, width=6.2cm]{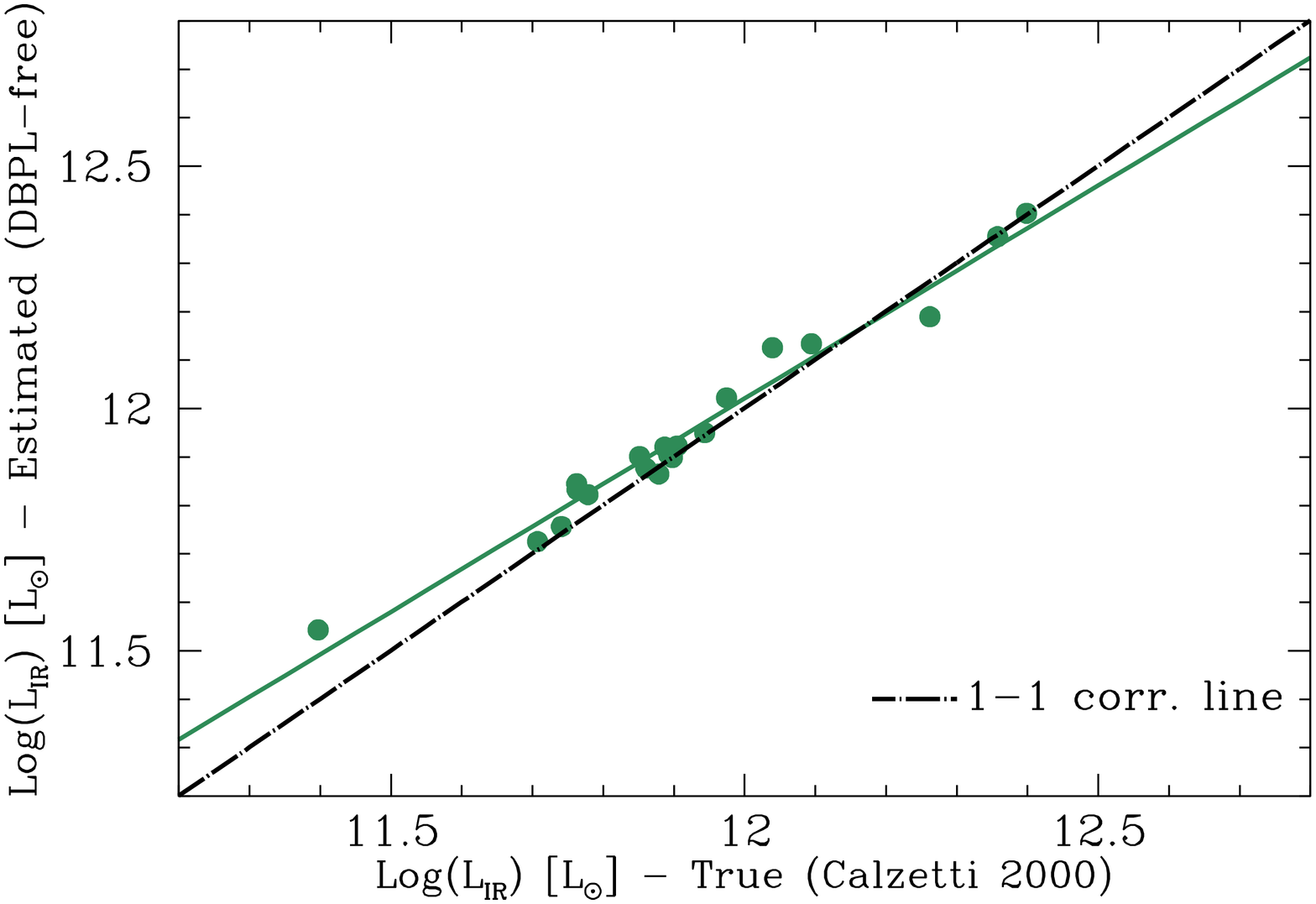}
\includegraphics[angle=-90, width=6.2cm]{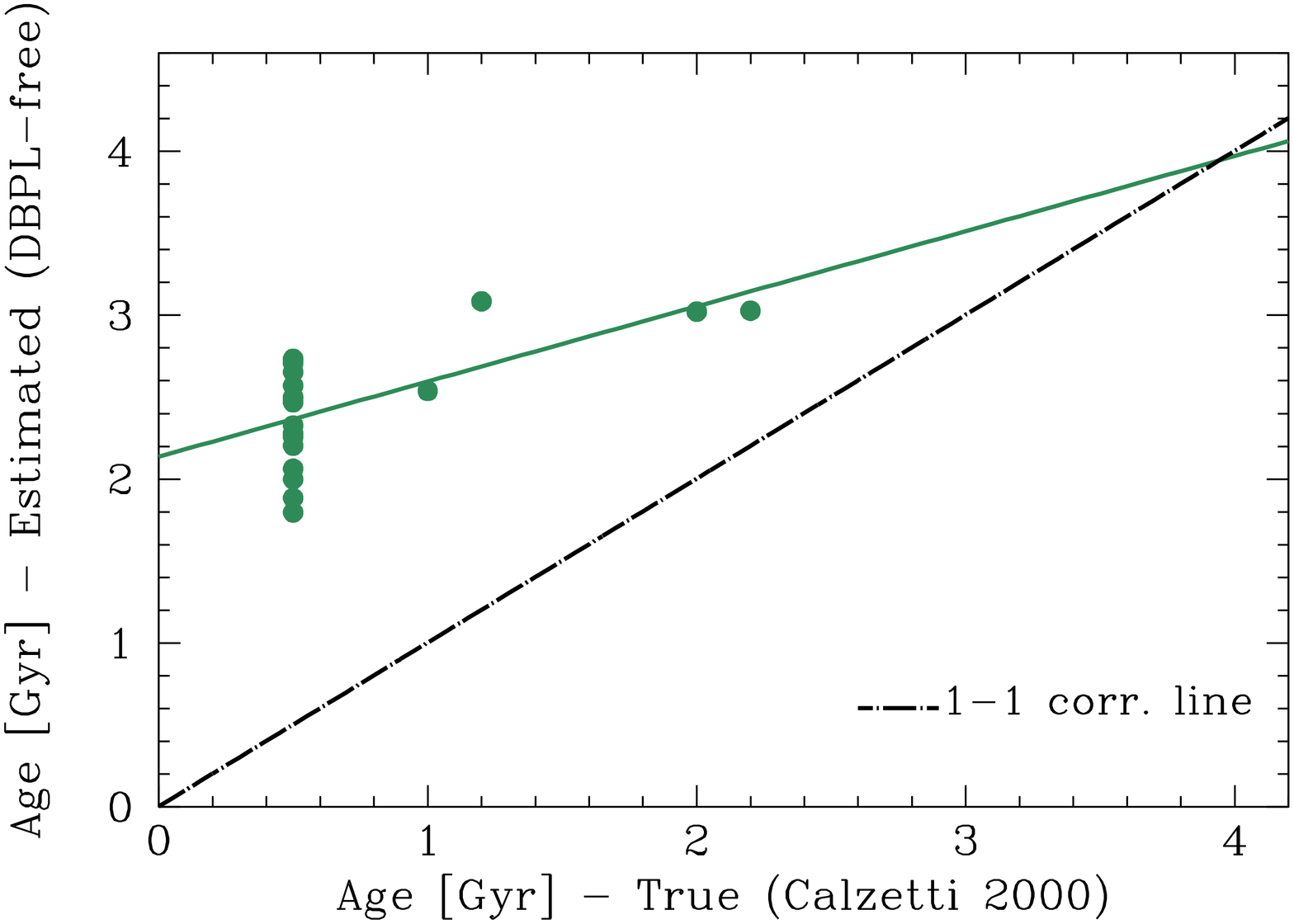}
\caption{Comparison between the predictions of the code under the DBPL-free configuration (y-axis) and the true values obtained by assuming the \citet{Calzetti2000} recipe (x-axis).  From top to bottom  the results relative to the SFR, M$_{*}$, L$_{\rm IR}$ and age are shown. The 1-1 correlation line is shown as black long dashed-dotted line in each panel.}
\label{dbpl-free-calz}
\end{figure}
CIGALE is then run on this artificial catalogue in order to compare the exact values of the physical parameters corresponding to the artificial SEDs to the parameters estimated by the code with      the PDF of each parameter.

The results are summarized in Figure~\ref{mock_analysis_fig}. The true value of the output parameter (artificial SED) on the x-axis is compared to the same parameter estimated by the code on the y-axis.  The 1-1 correlation line is shown as black long dashed-dotted line in each panel. From top left to bottom right the results for the star formation rate, stellar mass, total IR luminosity, age, e-folding timescale of delay-$\tau$ SFH, BC-to-ISM attenuation factor $f_{\rm att}$ and the slope of BC and ISM attenuation curves, are shown. The latter three quantities concern only our reference recipe, the DBPL-free. In each panel, the results from the mock analysis performed under the DBPL-free configuration (solid green line) are compared to those obtained with the standard recipes listed in Table~\ref{tab-1}. While the extensive physical quantities such as recent $SFR$, $M_{*}$, $L_{\rm IR}$  seem to be quite well constrained by our analysis, as long as the information provided by the full far-UV to sub-mm SED is available \citep[see e.g.,][]{Buat2014, Boquien2014}, the characteristic timescale of the assumed SFH appears unconstrained.
Constraining the detailed SFH of a galaxy from the observed SED is, indeed, a well known critical issue of spectral synthesis models \citep[see e.g.][]{Papovich2001, Daddi2004, Maraston2006, Santini2009, Wuyts2011, Pforr2012, Michalowski2012, Boquien2015, Ciesla2015}.

As will be discussed in detail in Section~\ref{discuss_att_curves}, one interesting aspect emerging from this mock analysis is that, although we seem to be able to constrain quite well the total amount of attenuation ($A_{\rm FUV}$ - central right panel, FUV corresponding to the FUV GALEX filter in the rest-frame of the galaxy) suffered by these galaxies, the exact shape of the attenuation curve appears to be more complex to constrain. This is related to the fact that the energy budget between the radiation absorbed in the UV and that one re-emitted in the IR does not depend significantly on the exact shape of the attenuation curve but rather on the total amount of attenuation extinguishing the UV-optical starlight.

For what concerns the shape of the dust attenuation curve under the DBPL-free configuration it is clear, from our mock analysis, that the slope of the power-law relative to the birth-cloud component cannot be constrained, at least with the information available on these objects. Conversely, the slope of the attenuation curve relative to the diffuse component (Fig.~\ref{mock_analysis_fig}-bottom right) appears to be well constrained with a Pearson's correlation coefficient for linear regression, between the true and estimated values, $\sim$ 0.80.

Another dust attenuation parameter which seems difficult to constrain from our analysis is the reduction factor for the attenuation of the old stellar population with respect to the younger one. We have performed several tests with CIGALE by considering this parameter both free and fixed but no relevant differences or dependencies in the results have been observed.
This specific parameter is known to be usually unconstrained by broad band photometry SED-fitting procedures \citep[see e.g.,][]{Noll2009,  Buat2011, Buat2012}.     Following \citet{Buat2012} a fixed value  of 0.5 is  adopted.

In the mock analysis described above the catalogue of artificial sources is created and analyzed, for each configuration, under the same assumptions for the input parameters. It does not consider cross-checks among different assumed parameter configurations. It can be interesting at this point to check the robustness of our parameter estimation also against variations of the assumed attenuation curve. An interesting comparison in this sense is the one involving the two ``extreme'' end recipes for the dust attenuation, namely our DBPL-free on the one side, and the \citet{Calzetti2000}, on the other.

In addition we thus consider the mock catalogue generated with the Calzetti~(2000) recipe and fit its artificial fluxes with the DBPL-free configuration. The results relative to  $SFR$, $M_{*}$, $L_{\rm IR}$ and galaxy age are shown, from top to bottom, in Figure~\ref{dbpl-free-calz}. The total IR luminosity appears, again, well constrained. The SFRs predicted by assuming a power-law free slope attenuation curve are systematically lower than the true values of \citet{Calzetti2000} by a factor $\sim$ 0.2 dex. These are consistent with the higher galaxy ages obtained, on average, with the DBPL prescription. Discrepancies in the predicted stellar masses are also observed with the DBPL-free configuration providing, on average, larger stellar masses than Calzetti. The impact of the different assumed dust attenuation prescriptions on the derived SFR and M$_{*}$ of IR selected galaxies and their consequences in a cosmological context will be discuss in detail in Section~\ref{sfr_mass_consequences}.

   We discuss now the results of the mock analysis concerning the amount of dust attenuation in  the UV and NIR bands which are  contributed by different stellar populations.

We have already seen that as long as the energy budget in the galaxy is preserved, the total amount of attenuation suffered by the galaxy is always well constrained. This is shown by Figure~\ref{fig_Anir_Afuv} where the results from our mock analysis concerning the estimates of the total dust attenuation in the FUV and NIR band are plotted color coded as a function of the different prescriptions considered. The  case where the catalogue of artificial sources is created and analyzed under the same assumptions for the input parameters is shown in the top panels of the figure.
\begin{figure*}
\centerline{
\includegraphics[angle=-90, width=8.2cm]{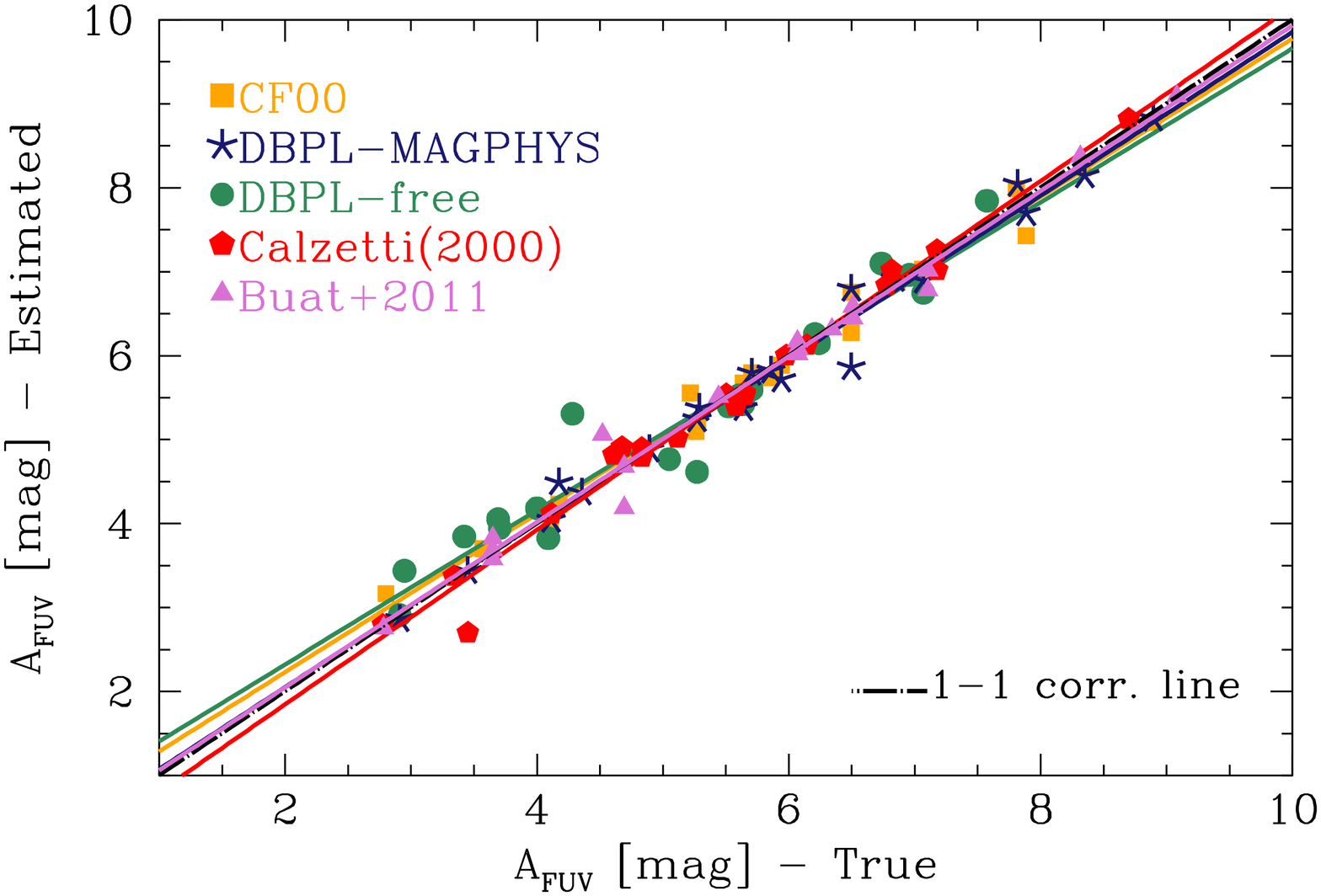}
\includegraphics[angle=-90, width=8.2cm]{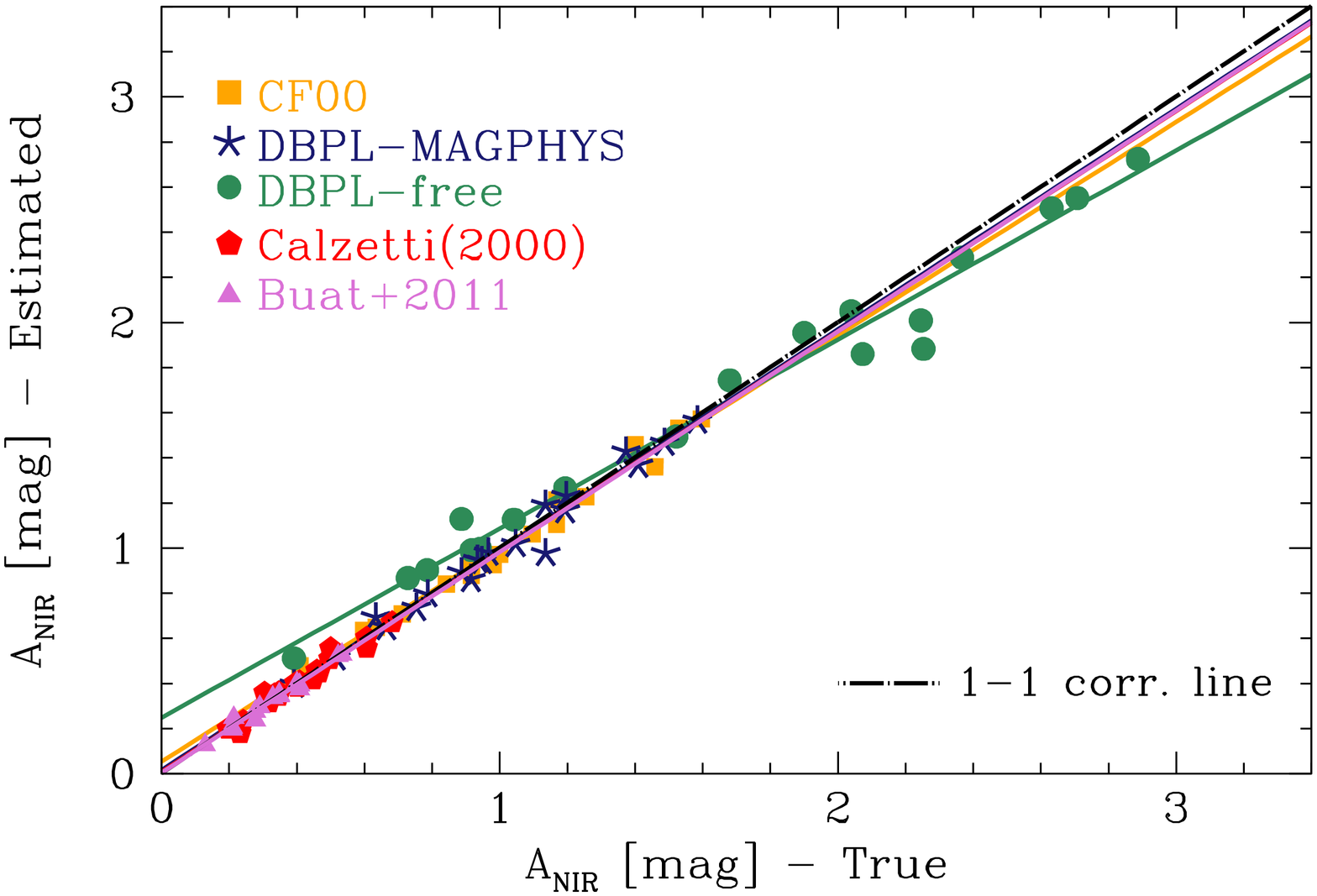}}
\centerline{
\includegraphics[angle=-90, width=8.2cm]{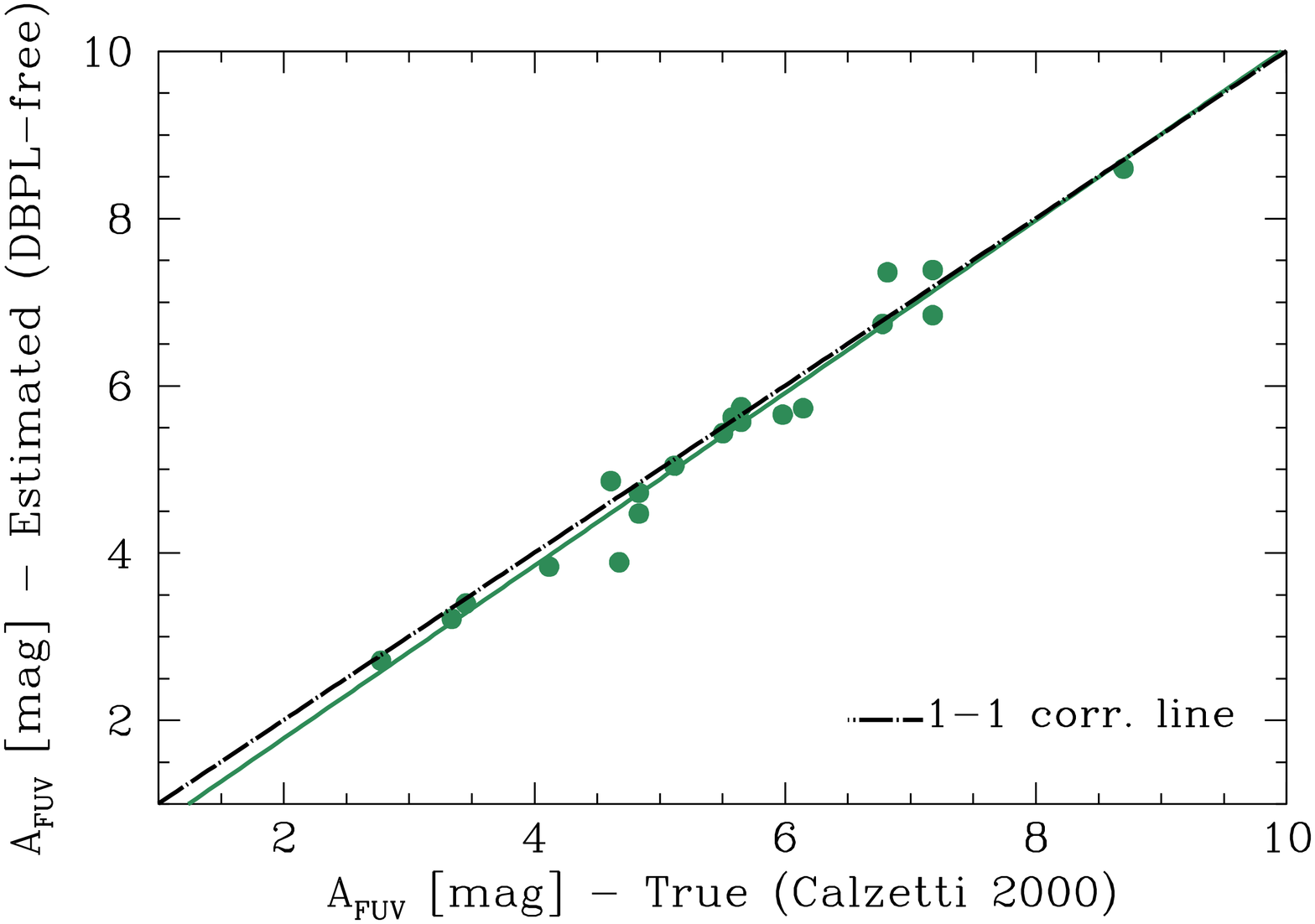}
\includegraphics[angle=-90, width=8.2cm]{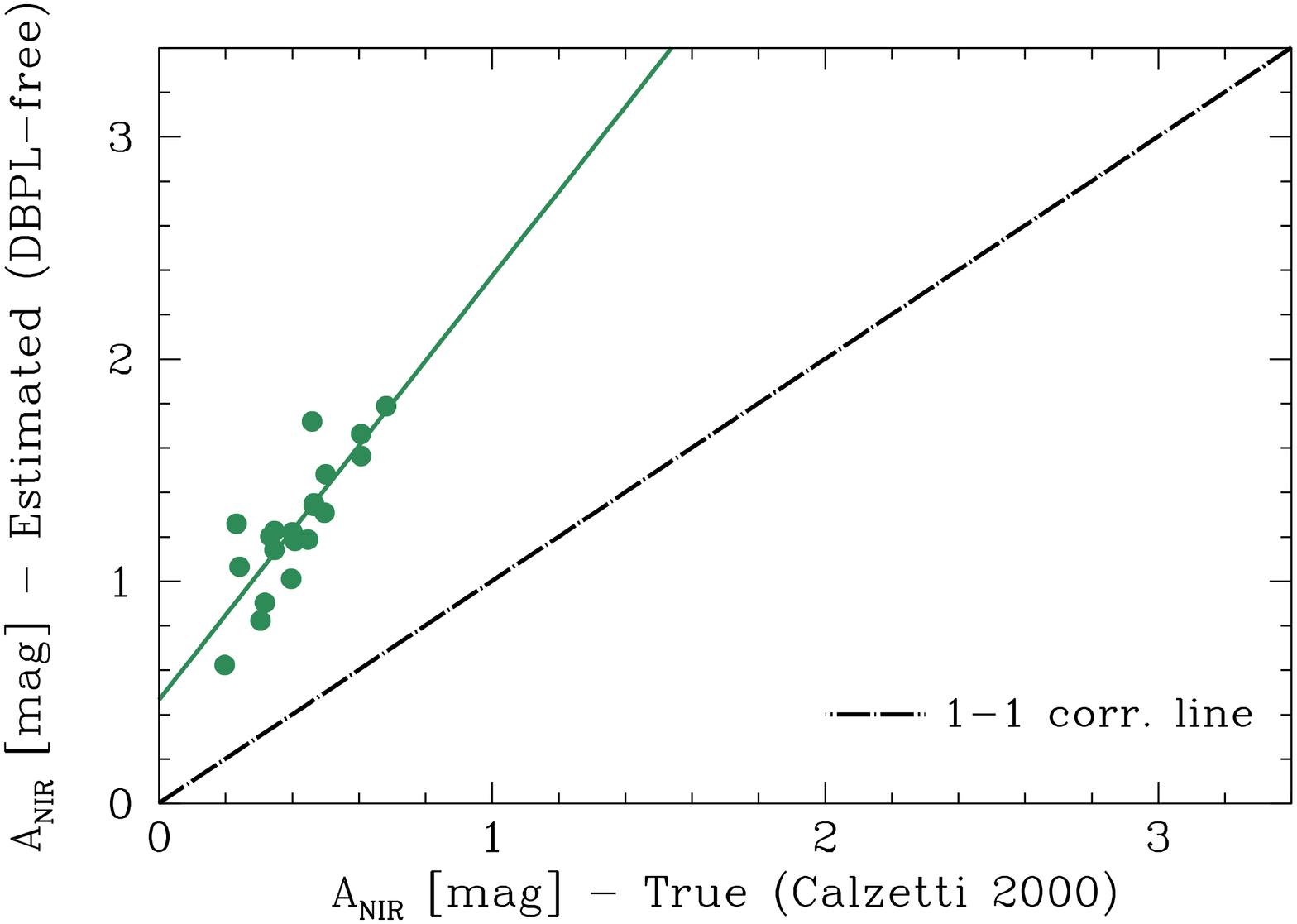}}
\caption{The results from our mock analysis concerning the estimates of the total dust attenuation in the FUV and NIR band are plotted color coded as a function of the different prescriptions considered. In the bottom panel we compare the true values corresponding to the assumption of a Calzetti dust attenuation curve to the mock analysis performed by assuming a DBPL-free model. }
\label{fig_Anir_Afuv}
\end{figure*}

In the bottom panels we compare, instead, the true values relative to the assumption of a Calzetti dust attenuation curve with the mock analysis performed by assuming a DBPL-free model. In this case, we observe that while the total amount of attenuation in the FUV remains well constrained, independently from the specific dust attenuation law assumed, the amount of attenuation predicted by the DBPL-free model at longer wavelengths appears to be larger, on average, than the true value given by Calzetti. This is directly linked to the intrinsic shape of the dust attenuation curve adopted, particularly at longer wavelengths. In fact, as previously shown in Figure~\ref{attenuation_curves_recipes_fig}, whenever the free power low slope equals the slope of CF00 model, ($\delta_{\rm ISM}$=-0.7), the UV-optical shape of Calzetti is well retrieved and so the  $A_{\rm FUV}$. However the same prescription tends to provide  grayer slopes than Calzetti at longer wavelengths thus bringing to a larger attenuation in the NIR.

\subsection{The dust attenuation curve of z$\sim$2 (U)LIRGs estimated with CIGALE}
\label{discuss_att_curves}

Figure \ref{dust_attenuation_curves} shows, for each of the z$\sim$2 (U)LIRGs in our sample, the normalized attenuation curve estimated with CIGALE under the assumption of our DBPL-free configuration (solid green line). The slope of the ISM component, as derived from the PDF analysis performed with CIGALE, is also highlighted in each panel. The value of the slope of the BC component is omitted here because, as we have seen in Figure~\ref{mock_analysis_fig}, it is not constrained by our broad band analysis. The DBPL-free attenuation curves are then compared, in the same figure, to those derived with standard recipes, namely, MAGPHYS-like in blue, \citealt{Calzetti2000} in red and \citealt{Buat2011} in gray.\footnote{Due to the differential attenuations, the curves plotted are not exactly the \citet{Calzetti2000} or \citet{Buat2011} curves.
The figure clearly shows a global flattening up to longer wavelengths of the effective attenuation curves computed using double power-law recipes compared to those derived using Calzetti-like prescriptions. It is therefore clear that a larger amount of attenuation in the NIR is implied, for our IR luminous galaxies, when using either MAGPHYS-like or DBPL-free recipes.

An example of a best-fit SED obtained with CIGALE for a representative objects, \textit{U4451}, is given in Figure~\ref{sed-nir-effect} which will be discussed in Section~\ref{stellar_mass_effect}. }
\begin{figure*}
\centering
\includegraphics[width=17.cm]{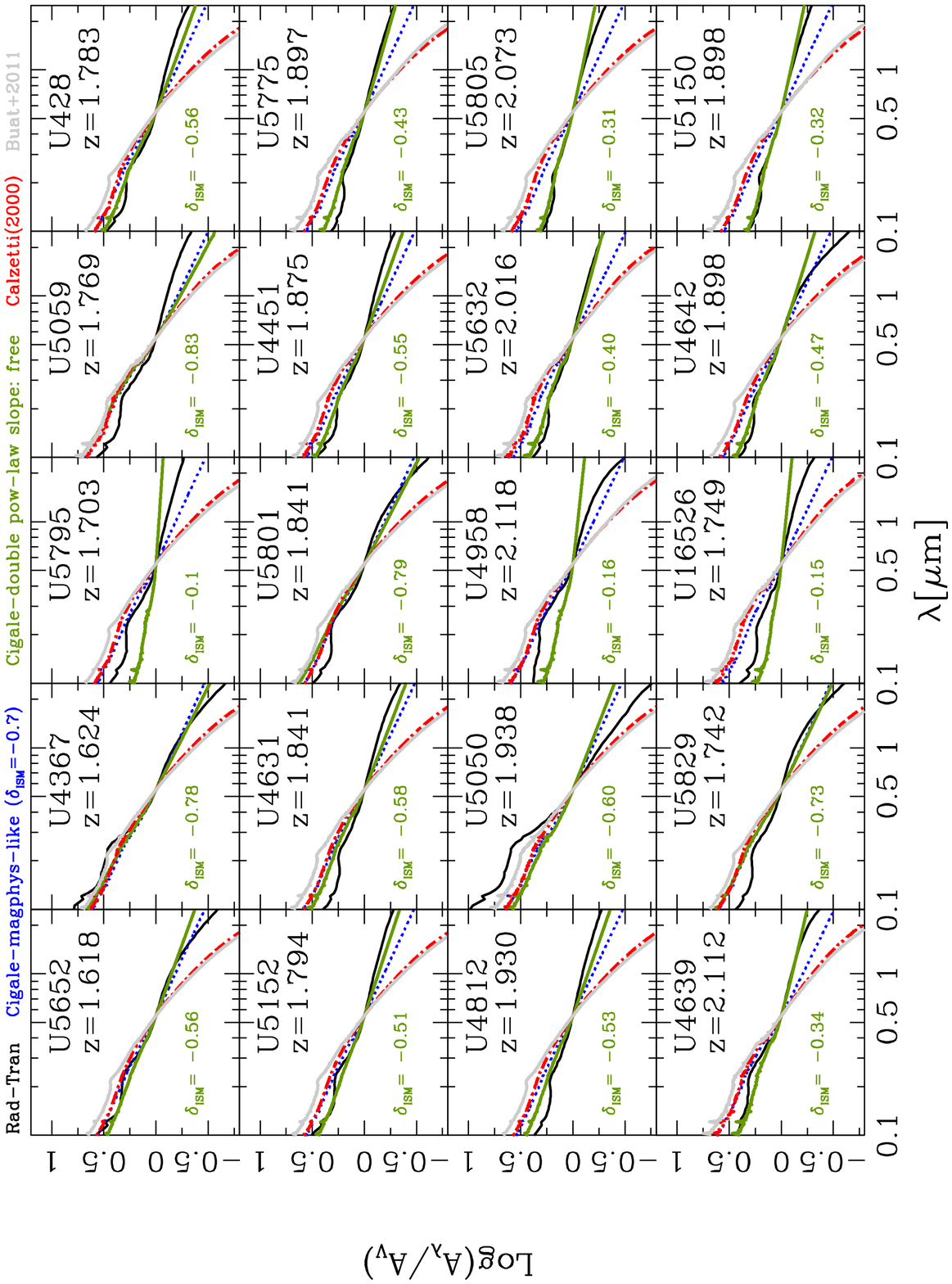}
\caption{ Attenuation Curves of z$\sim$ 2 (U)LIRGs. The attenuation curves are color-coded as a function of the different adopted prescriptions for the dust attenuation, as reported on the top of the figure. See text for details. }
\label{dust_attenuation_curves}
\end{figure*}

\subsection{Comparison with RT-derived attenuation curves}
\label{rt_modelling}

We present in this section the  attenuation curves based on radiative transfer and used for the purpose of comparison with the results found in this work.

In previous works \citep[][]{LoFaro2013, LoFaro2015} we analyzed the stellar masses and SFRs of diverse galaxy samples,
among which the same $z\sim2$ (U)LIRGs considered in this work. The analysis was performed by fitting the observed UV to sub-mm and radio SEDs of galaxies with the spectral synthesis and radiative transfer code GRASIL \citep[][]{Silva1998, Silva2011, Vega2008}. It is therefore interesting to compare and discuss the attenuation curves obtained with independent and very different modelling for the same data. In RT models, in fact, the attenuation curve is not an input parameter as in physically motivated SED-fitting procedures, but it results from the complex interplay among SFH, dust properties and star-dust geometry, as detailed below.

We provide here a short summary of the main features and techniques used in these previous works, and defer to the original papers for details on GRASIL, and in particular to the Lo Faro et al. papers for the specific approach and details on the model parameters used for the fits.

GRASIL is a spectral-synthesis code for the SED of galaxies which computes the radiative transfer of the stellar radiation through a two-phase dusty medium. A dense phase, representing molecular clouds associated with newly born stars,
is distributed within a diffuse medium, the cirrus, associated with more evolved stars.
The clumping of both young stars and dust within the diffuse medium gives rise to an age-dependent dust attenuation
where younger stars are more extinguished than older stars.
A wide spread in the shapes of the output attenuation curves can be obtained, including the Calzetti (2000) curve,
depending on the input SFHs and age-dependent star-dust distribution \citep[e.g.][]{Granato2000, Panuzzo2007, Fontanot2009b}.

The input SFHs used in Lo Faro et al., were computed with a standard chemical evolution code and Salpeter IMF and
have shapes consistent with the delay-$\tau$ models. The best fits are characterized by very short
$\tau$ for half of the sample, and by more steadily evolving SFHs for the other half. No burst
was required for these high-z (U)LIRGs (consistently with their MS nature). The best fit was searched within a large model library including $\sim 10^6$ models, each one corresponding to a different combination of parameters among which the geometrical scale-lengths, the fraction of gas in MCs with respect to the diffuse component, the typical disruption timescale of giant MCs and galaxy age. A description of these model parameters and their range of values can be found in Table  1 of \citet{LoFaro2013}.
We stress that we rely here on the assumptions and specific parameter configurations adopted in \citet{LoFaro2013,LoFaro2015}. We do not discuss here the effects of different geometries or dust parameters on the RT-derived dust attenuation curves. A discussion about the implications of a disk versus spheroidal geometry on the observed SED of our
sources is already partly dealt with in \citet{LoFaro2015} and will be treated in more detail in a future work.

Figure~\ref{grasil_attenuation_curves} shows the attenuation curves (solid black lines) obtained with GRASIL for the sample of z$\sim$2 (U)LIRGs.
The curve obtained by averaging over the entire sample is also shown as solid green thick line. Compared to the \citet{Calzetti2000} attenuation law (red long-dashed line), they appear on average  grayer over the entire wavelength range.
The same RT-based attenuation curves are compared to those estimated with CIGALE in Figure~\ref{dust_attenuation_curves}.

The double power law recipes provide attenuation curves in most cases consistent with those derived from radiative transfer modelling.
    For four objects, {\it U5795, U5050, U4958, U16526}, the curves obtained with the DBPL-free fit are much grayer than those yielded by the RT fit. Two of them ({\it U5795} and {\it U16526}) have  the largest $\chi^{2}_{\nu}$ (5.7 and 4.7 respectively with the mean $\chi^{2}_{\nu}$ over the entire sample being $\sim$2.2). The SEDs of {\it U5050} and {\it U4958} are well fitted ($\chi^{2}_{\nu}$ $\simeq$ 1.5), the curve found for {\it U5050} is only slighly flatter than the original CF00 recipe (-0.6 against -0.7). In one case, {\it U5059},  the DBPL fit leads to an attenuation curve steeper than the RT fit but not very different from the CF00 recipe ($\delta_{\rm ISM}=-0.83$, with $\chi^{2}_{\nu}$ = 1.7).
\begin{figure}
\centering
\includegraphics[width=8.6cm]{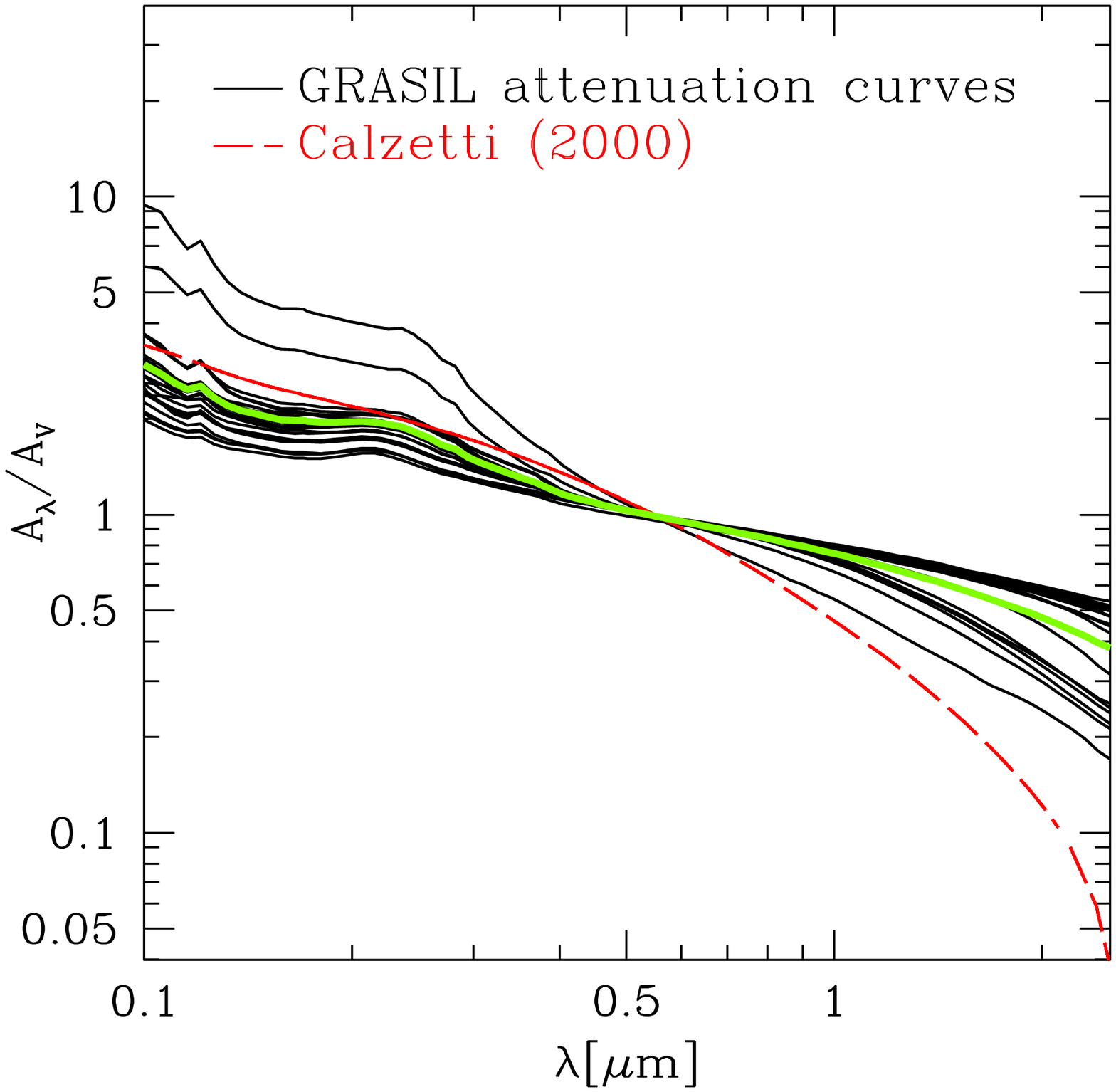}
\caption{Effective attenuation curves (solid black lines) obtained with GRASIL for the reference sample of (U)LIRGs at $z\sim2$ (Lo Faro PhD thesis). The curve obtained averaging over the entire sample is also shown as solid green thick line. Compared to the classical \citet{Calzetti2000} attenuation curve (red long-dashed line), they appear on average  grayer over the entire wavelength range, particularly in the NIR.
}
\label{grasil_attenuation_curves}
\end{figure}

An interesting aspect emerging from this comparison is the wide range of attenuation curve shapes allowed by the RT modelling and DBPL-free recipe with respect to assuming a fixed slope for the attenuation curve. This is shown in Figure~\ref{att_curve_mean} (left panel), where the V-band normalized attenuation curves derived with CIGALE (MAGPHYS-like in blue in the top right panel, the two Calzetti-like recipes in the bottom left panel and DBPL-free in green in the bottom right panel),
are compared to those derived with the RT code (top left in black).

\begin{figure*}
\centerline{
\includegraphics[width=8.6cm]{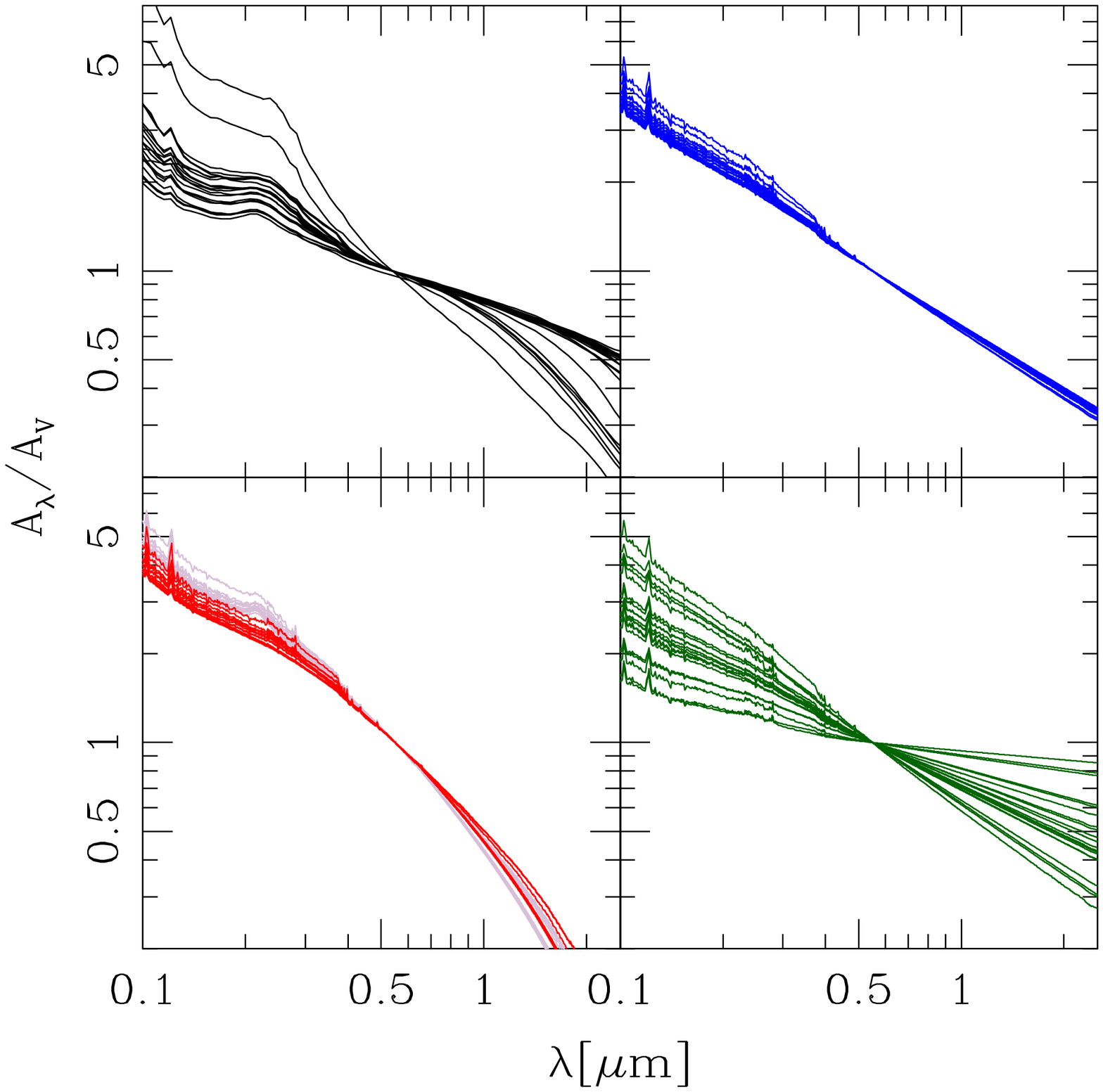}
\includegraphics[width=8.6cm]{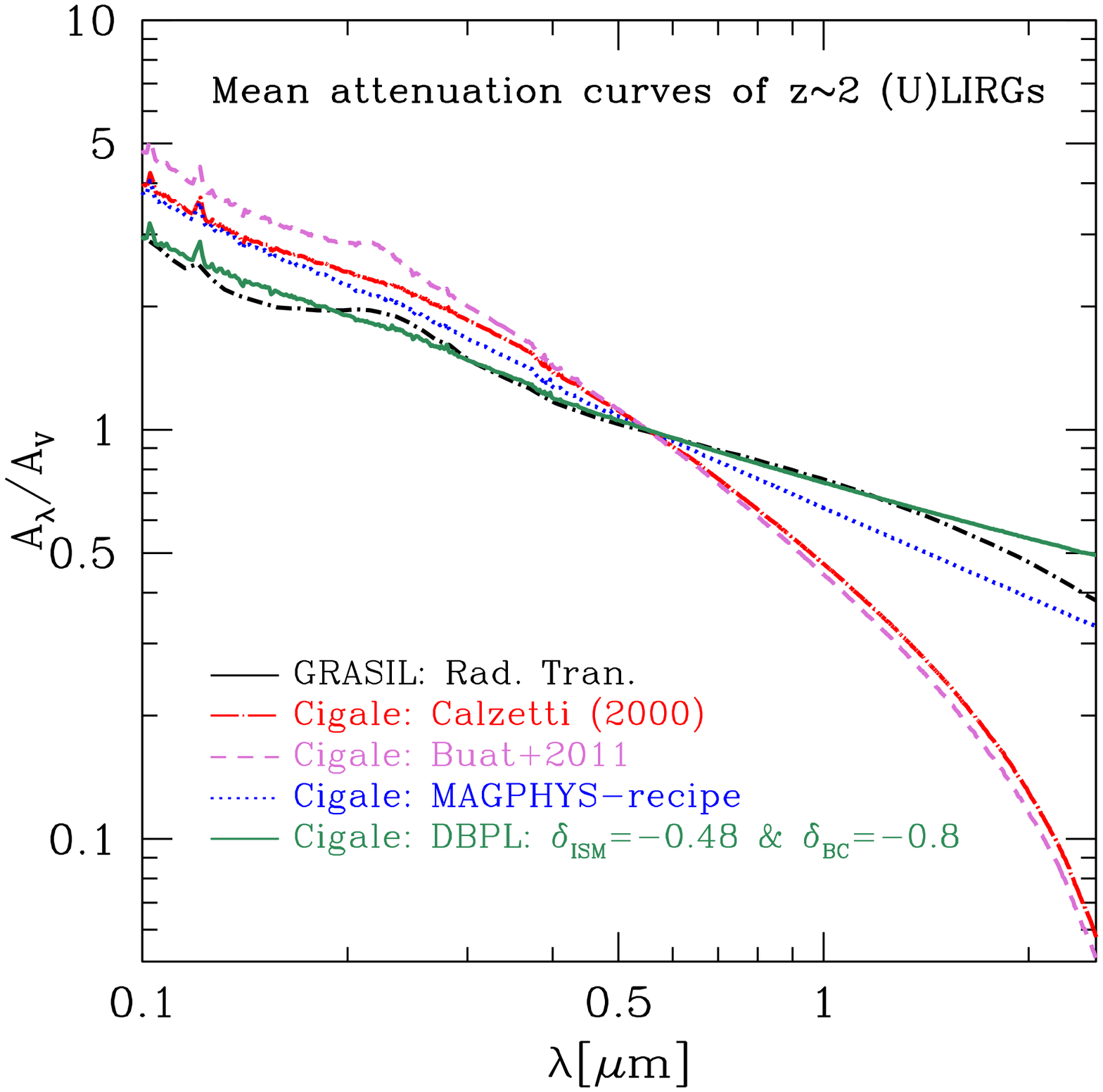}}
\caption{Left: the variability of the normalized effective attenuation curves derived from radiative transfer modelling (top-left in black) is compared to that one of the curves computed with CIGALE under the different assumptions for the dust attenuation. From top-left to bottom-right are shown, MAGPHYS in blue, \citealt{Calzetti2000} in red and \citealt{Buat2011} in gray and our DBPL-free recipe in green. Right: Mean attenuation curves, averaged over the entire sample, of z$\sim$2 (U)LIRGs color coded as a function of the different empirical recipe adopted and compared to the mean attenuation curve derived from radiative transfer computations (black dot- dashed line). \citet{Calzetti2000} is shown as red dot-long dashed line, \citet{Buat2011} as magenta dashed line,MAGPHYS  as blue dotted line and our DBPL-free as solid green line. The slope of the ISM component of the free power law recipe, as derived from the PDF analysis performed with CIGALE, is also highlighted in each panel. The flattening at longer wavelengths of the free slope power-law recipe appears to be quite consistent with what predicted from radiative transfer modelling.}
\label{att_curve_mean}
\end{figure*}
As anticipated above, the larger spread characterizing the attenuation curves derived from RT models can be explained as arising from a complex combination of star formation histories and age-dependent relative distribution of stars and dust, including the clumping of both components \citep[see e.g.,][]{Granato2000, Panuzzo2007, Fontanot2009b}. The relative importance of these ingredients is a function of the characteristics and evolutionary status of the galaxy.
    In very active systems  a large fraction of the rest-frame UV-optical starlight is produced by stars embedded in
molecular clouds while older stars, mainly emitting in the optical and NIR, suffer a smaller effect from the diffuse medium. These arguments were used by \citet{Granato2000} and \citet{Panuzzo2007} to explain the differences  between the shallow attenuation in the UV of starburst galaxies (i.e. the Calzetti law) and the MW average extinction law, even by adopting a fixed, MW-type dust model.
Similar conclusions have been reached by \citet{CharlotFall00, Pierini2004, Tuffs2004, Inoue2005, Inoue2006} and more recently by \citet{Chevallard2013}.
In particular, \citet{Chevallard2013} compared the results from different radiative transfer models including GRASIL, \citet{Pierini2004}, \citet{Tuffs2004} and \citet{Jonsson2010} and found all of them to accordingly predict a quasi-universal
relation between the slope of the attenuation curve at any wavelength and the V-band attenuation optical depth in the
diffuse ISM at all galaxy inclinations.
All the models yield a flattening of the dust attenuation curve at increasing optical depths.

The shape and variation of the attenuation curve can be quantified also in terms of the total-to-selective extinction ratio in the V-band, $R_{\rm V} = A_{\rm V}/E(B-V)$.
In star-dust configurations with clumping of both components and connection between stars of different ages with different dusty environments we do expect the path length through the dust to the observer to be different for each star. Therefore the light from each star will experience a different amount of attenuation, yielding an attenuation curve for the entire galaxy
which can be significantly different from the input extinction curve of the dust grains, with a value for $R_{\rm V}$ that can significantly depart from the value of 3.1 characteristic of the average Milky Way extinction curve \citep[see e.g.,][]{Pierini2004, GonzalezPerez2013, Mitchell2013, Scicluna2015}.
Variations of $R_{\rm V}$ as a function of V-band optical depths were indeed found by \citet{Pierini2004} for both homogeneous and clumpy media with $R_{\rm V}$ increasing at larger optical depths. The non-linear dependence of $R_{\rm V}$ on the total amount of dust was identified as the main cause of the flattening observed at large optical depths. More recently \citet{Scicluna2015} using a Monte Carlo radiative transfer code also found clumping dust distributions to be able to reproduce attenuation curves with arbitrary $R_{\rm V}$ .

    In the \citet{Calzetti2000} recipe, the local UV-bright starburst galaxies used for the calibration, provide a relation where the ratio $R_{\rm V} = A_{\rm V}/E(B-V)$ is fixed to $4.05$. For the MAGPHYS-like recipe the fixed input slopes for the ISM and BC component of the MAGPHYS-like recipe, ($\delta_{\rm ISM}$ =-0.7 and $\delta_{\rm BC}$ = -1.3) lead to a  total-to-selective extinction ratio  equal $\sim 5.8$.
This larger value illustrates the flatter shape of MAGPHYS attenuation curve compared to \citet{Calzetti2000}. As a consequence the MAPHYS-like recipe is in better agreement with the results of RT modelling than  \citet{Calzetti2000}. In this work  we implement  attenuation curve recipes where the shape is left free to vary, mimicking, implicitly, the effect of different geometrical configurations and situations. This is one of the main advantages of working with physically motivated SED-fitting codes and it also explains why our flexible DBPL-free recipe is able to reproduce the large variations observed in the radiative transfer models.

The right panel of Figure~\ref{att_curve_mean} finally shows the mean attenuation curves of our $z \sim 2$ (U)LIRGs averaged over the entire sample, color coded according to the specific recipe considered. An overall flattening from far-UV to NIR, compared to the Calzetti-like recipes, is accordingly yielded by both the DBPL-free (solid green line) and RT fits (black dot-dashed line).

\subsection{IRX-$\beta$ as ``a posteriori'' diagnostic to constrain the UV shape of the assumed attenuation law}
\label{IRX_beta}
Given its ascertained sensitivity to the assumed attenuation curve we use here the IRX-$\beta$ diagram \citep[][]{Meurer1999} as a ``posteriori'' diagnostic to constrain the UV shape of the dust attenuation curve of our high-$z$ IR luminous sources.

Since the work by \citet{Meurer1999}, on local UV-bright starburst galaxies, the IRX-$\beta$ relation has been widely used at all redshifts to estimate the total amount of attenuation of distant galaxies in the FUV. Deviations from this relation have been interpreted as due to either different attenuation laws or SFHs, or both \citep[see e.g.][]{Kong2004, Panuzzo2007, Conroy2010, Forrest2016, Salmon2015}.
Local LIRGs and ULIRGs, for example, are  found to fall above the locus defined by Meurer relation \citep[e.g.,][]{Goldader2002, Howell2010}. These galaxies are usually characterized by larger IR excess for their UV slopes and clumping geometries responsible for grayer attenuation curves than \citet{Calzetti2000}. Steeper curves like SMC, are instead  used to fit more normal star forming sources lying below that relation and characterized by low IR excesses for their UV colours. In such systems, the UV colour, seems to be very sensitive to influences from both stellar populations and dust geometry variations \citep[e.g.,][]{Kong2004, Calzetti2005, Boquien2009, Hao2011}.

In a recent work \citet{Salmon2015} analyzed the UV-shape of the dust attenuation curve of a sample of IR luminous sources at redshift 1.5-3 by considering several different recipes for the dust attenuation curve including Calzetti and modified Calzetti \citep[][]{Noll2009} laws and a steeper SMC curve. They found galaxies with high color excess to have shallower, starburst-like attenuation, while those characterized by low color excess to have steeper, SMC-like attenuation.
Before discussing the position of our galaxies in the IRX-$\beta$ diagram and its interpretation in terms of the UV shape of the measured attenuation curve, we provide in the next subsections few details about how the IRX ratio and UV-spectral slope have been computed in this work.
\begin{figure*}
\centerline{
\includegraphics[angle=-90, width=9.5cm]{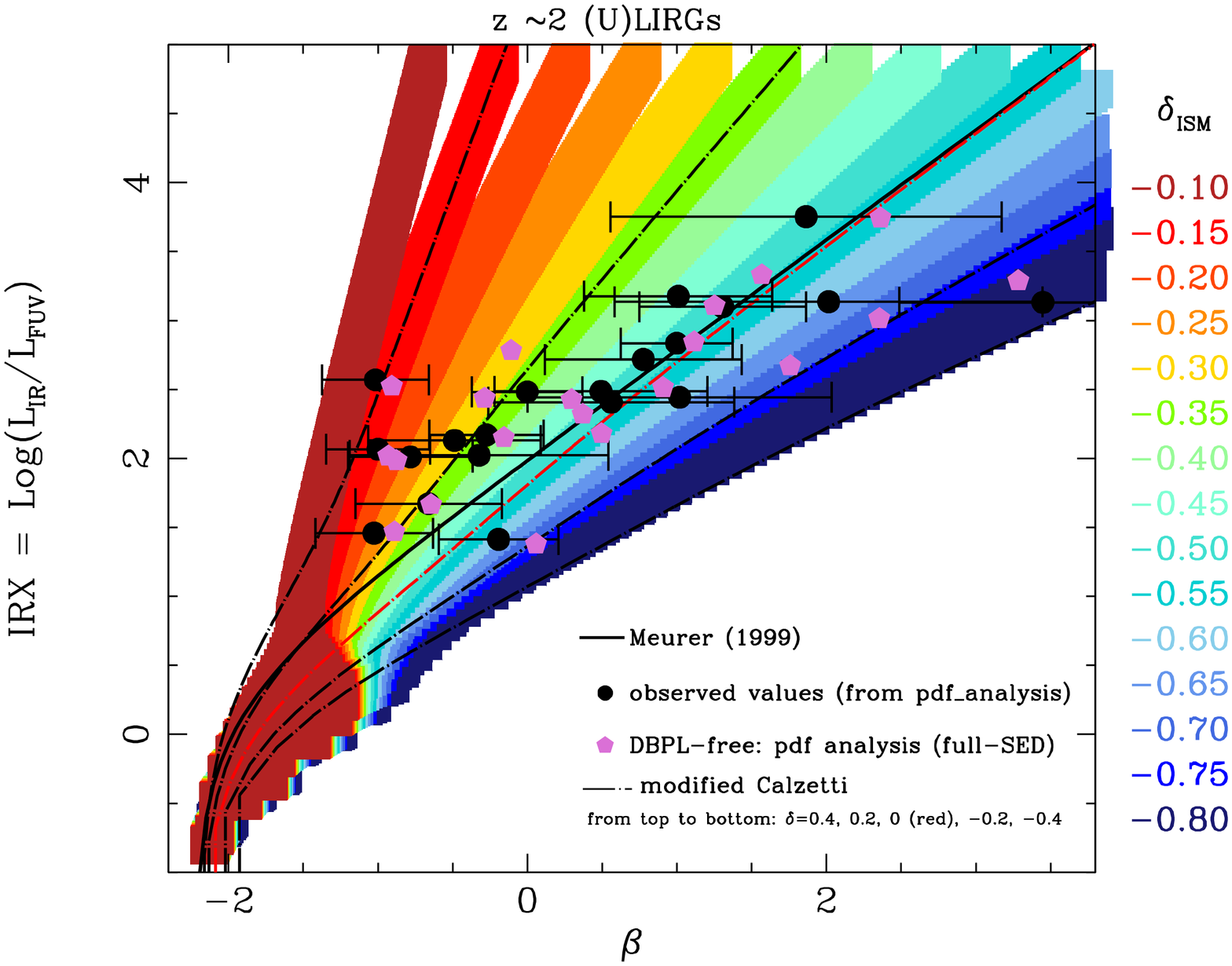}
\includegraphics[angle=-90, width=9.5cm]{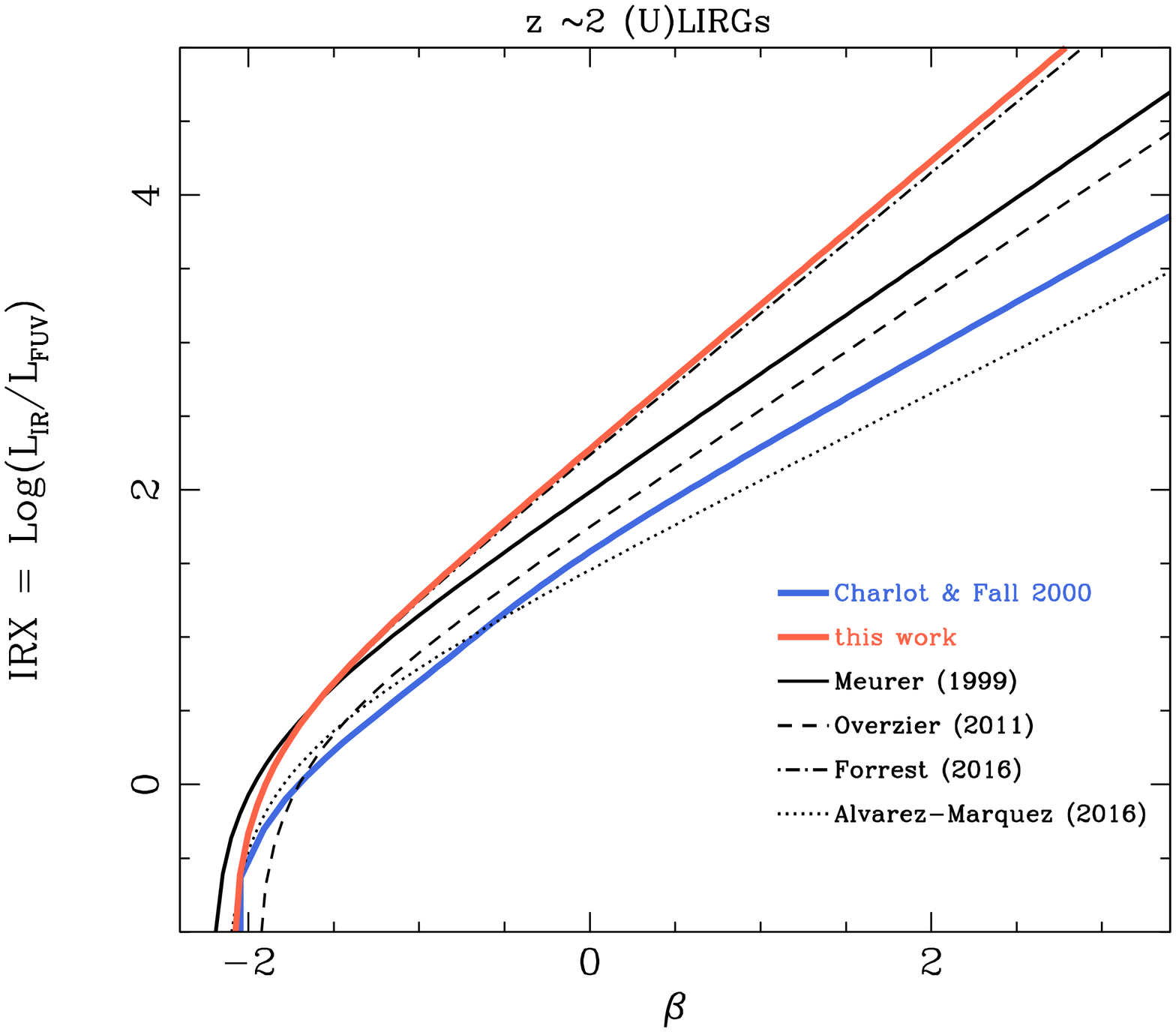}}
\caption{Left: observed estimates for the IRX-$\beta$ of our z$\sim$2 (U)LIRGs (filled black circles) compared to the predicted relations obtained through simulations color coded from blue to red as a function of increasing ISM slopes from -0.8  to -0.1. The thickness of the different colored stripes reflects the effect due to the assumed SFH, stellar population properties and different E(B-V)/A$_{\rm V}$ considered (see Tab.~\ref{tab-1} for reference). These are also compared to the reference Meurer relation shown as solid black line and the modified Calzetti recipes, used in \citet{Salmon2015}, with slopes ranging, from top to bottom, from +0.4 to -0.4. The red dot-long dashed line being that one corresponding to the assumption of a \citet{Calzetti2000} attenuation law ($\delta=0$ in the modified Calzetti recipe). The latter are used here as reference to show that in terms of UV-slope we cover the same range of values and they have been computed by assuming our reference SFH, namely, delay-tau model with input parameters as defined in Tab.~\ref{tab-1}. The magenta pentagons correspond to the estimates obtained through full SED-fitting under the assumption of DBPL-free model. Right: the IRX-$\beta$ relation (red solid line) obtained by fitting the observed estimates from our z$\sim$2 (U)LIRGs, is compared to that one obtained from simulations by assuming the CF00 model (blue solid line) and several reference relations from literature. Our relations appears to be in very good agreement with that one recently derived by \citet{Forrest2016}.}
\label{IRX-beta-ulirgs}
\end{figure*}

\subsubsection{Computation of the UV-spectral slope $\beta$ and rest-frame FUV luminosity}
\label{beta}

As discussed above the UV-spectral slope can be an important observational tool, particularly at high redshift, for its relatively ease of measurement.
Originally $\beta$ was determined from UV-continuum spectroscopy \citep[][]{Calzetti1994}, however nowadays it is mostly computed from the UV colors provided by broad band photometry. The classical method implemented consists in a power-law fit to the observed photometric bands. This method has been shown to provide redder $\beta$ compared to the true values derived from stellar population models using the spectral windows defined by \citet{Calzetti1994}, with a systematic offset at all redshifts ranging between 0.1 up to 0.5 \citep[][]{Salmon2015}.

Independent tests using simulations and stacked samples of high-$z$ LBGs have been also performed (Alvarez-Marquez et al. in prep.) providing similar systematic off-sets when comparing $\beta$ values derived from SED-fitting and simple power-law fitting.

Here we use the method described in \citet{Finkelstein2012}, directly included in CIGALE, where $\beta$ is estimated from the best-fit to the UV-to-optical observed SED using the UV spectral windows defined by \citet{Calzetti1994}\footnote{We use  the spectral windows defined by \citet{Meurer1999} from 1268 to 1950 \AA}. This method has been shown to better recover the true values of $\beta$ with no clear systematics and a scatter around $\pm 0.1$ up to z$\sim$4 \citep[][]{Salmon2015}.   as it is used here is reference for local starburst galaxies.

Given the  redshifts (varying between 1.618 and 2.118) of our sources we have selected a subsample of about 13 sources with at least 3 photometric bands in the rest-frame UV-optical in order to check any possible effect in the estimated $\beta$ due to the presence of a possible UV bump. We compared the results obtained by including the rest-frame band (for galaxies at this redshift being the observed V-band filter) where the absorption feature can possibly fall with those obtained removing this band. We also compared colors for the two runs and did not find any significant effect concluding that for these sources the UV-bump does not seem to affect our estimates of $\beta$.

We also found, in agreement with \citet{Salmon2015}, our estimates of $\beta$ to be quite independent of the specific attenuation law assumed. This is due to the fact that the best-fit UV-optical SED always provides a close match to the UV colors as long as the assumed dust law does not have extreme features.
Here we use as a reference the $\beta$ values obtained through our own DBPL-free recipe.

Finally the rest-frame FUV luminosity used to compute IRX is obtained from the best-fit SED by fitting only the UV-optical bands in order to retrieve the best observed estimate of this quantity for our sources.

\subsubsection{The observed IRX-$\beta$ plot for our z$\sim$2 (U)LIRGs}

Figure \ref{IRX-beta-ulirgs} (left) shows the observed estimates for the IRX-$\beta$ of our z$\sim$2 (U)LIRGs (filled black circles) compared to the predicted relations obtained through simulations color coded from blue to red as a function of increasing ISM slopes from -0.8 (steeper slope than Calzetti) to -0.1 (much grayer slope than Calzetti). The thickness of the different colored stripes reflects the effect due to the assumed SFH, stellar population properties and different $E(B-V)$ or $A_{\rm V}$ considered (see Table~\ref{tab-1} for reference). These are also compared to the reference Meurer relation shown as solid black line and the modified Calzetti recipes, used in \citet{Salmon2015}, with slopes ranging, from top to bottom, from +0.4 to -0.4. The red dot-long dashed line being that one corresponding to the assumption of a \citet{Calzetti2000} attenuation law ($\delta=0$ in the modified Calzetti recipe). The latter are used here as reference to show that in terms of UV-shapes we cover the same range of values and they have been computed by assuming our reference SFH, namely, delay-tau model with input parameters as defined in Tab.~\ref{tab-1}.

The observed $\beta$ obtained through the method specified above and represented by the filled black circles are then compared also with the results from full-SED fitting (magenta filled pentagons). We find good agreement between the observed estimates and those derived with SED-fitting using the DBPL-free attenuation recipe. This reflects on the fact that the positions of our sources in the IRX-$\beta$ plot are fully consistent with the value of the ISM slope derived from Bayesian analysis. Moreover the figure clearly shows that grayer slopes than \citet{Calzetti2000} are needed in order to reproduce the observed IRX-$\beta$ of the objects in our sample characterized by bluer $\beta$ values (i.e. those falling roughly in the dark green-to-red area).
We consider this as an indication of flattening of the attenuation curve in the UV as observed in figure~\ref{att_curve_mean}. Figure~\ref{IRX-beta-ulirgs} (right) shows the mean IRX-$\beta$ relation derived for our z$\sim$2 (U)LIRGs (as red thick solid line) compared to some reference relations including \citet{Meurer1999} (black solid line), \citet{Overzier2011} (black dashed line), \citet{Forrest2016} (dot-log dashed line), \citet{AlvarezMarquez2016} (dotted line) and the same  relation obtained under \citet{CharlotFall00} prescriptions. The observed flattening of our attenuation curves at all wavelengths is reflected in an IRX-$\beta$ relation which is systematically above the local relations. It also appears to be in a very good agreement with the relation derived by \citet{Forrest2016} for dusty star forming composite SEDs selected from a sample of  $\sim$ 4000 K-band selected sources in the ZFOURGE survey.

Of course the IRX-$\beta$ plot does not provide any information about the specific shape of the dust attenuation curve at longer wavelengths and our conclusions in this case restrict only to the UV shape. This means that in the UV, either a power-law attenuation curve with slopes ranging between -0.7 and -0.1 (or a modified Calzetti recipe with grayer slope) would be good in reproducing the observed IRX-$\beta$ values of our galaxies.

\subsection{Constraining the NIR shape of the dust attenuation curve}
\label{bayes_fac_def}

\begin{figure}
\centering
\includegraphics[width=8.5cm]{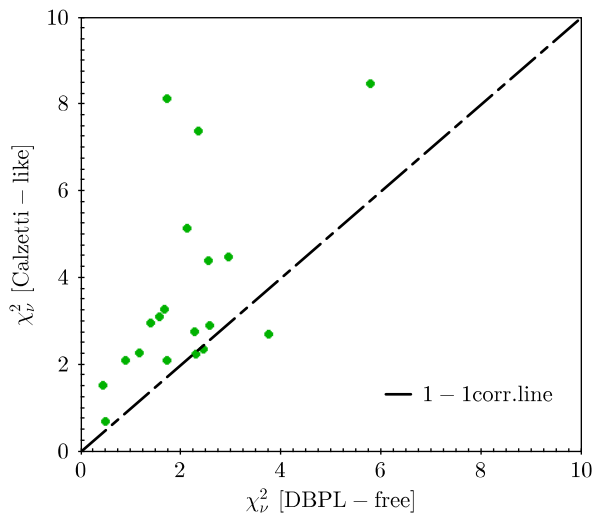}
\caption{Comparison of $\chi^2_{\rm \nu}$ values for z$\sim$2 (U)LIRGs obtained with the   DBPL-free recipe (x-axis)  and the modified  Calzetti recipe (y-axis). See text for further details.}
\label{chi2}
\end{figure}

     We want to understand whether we need or not, for a given range of UV slopes, a flattening in the NIR. At this aim we  compare the ability of the modified Calzetti recipe and the DBPL-free model, whose shapes strongly differ in the NIR, to reproduce our data. In order to isolate the effect of the assumed attenuation curve when comparing the two different set of models, we adopt exactly the same configuration for all the other parameters, namely the SFH and IR emission templates and we require  to cover exactly the same range of UV slopes for the attenuation curve (i.e. grayer to steeper than Calzetti, see figure~\ref{attenuation_curves_recipes_fig} for details). The four reference values for the ISM slope used in DBPL-free recipe (-0.7, -0.5, -0.3, -0.1) are compared to the corresponding four values in the modified Calzetti recipe (0., 0.2, 0.4, 0.6 with the 0 corresponding to the reference \citealt{Calzetti2000} value). In order to have the same number of parameters the slope of the BC component has also been fixed to the value of -0.7 \citep[][]{CharlotFall00}. Within this configuration the only difference between the two attenuation models pertains to the long wavelength range of the attenuation curve with the DBPL-free providing systematically grayer curves than modified Calzetti. We then run CIGALE under the two configurations and  compared the resulting $\chi^2_{\nu}$ in  figure~\ref{chi2}. Only one object, {\it U4642}, appears to be best fitted by the modified Calzetti recipe. Interestingly,   the RT modelling also gives a steeper attenuation curve than our DBPL-free result in the NIR for this galaxy (Figure~\ref{attenuation_curves_recipes_fig}).

\section{Dust attenuation configuration for LIRGs and (U)LIRGs at lower redshift}
\label{low_redshift_checks}

By fitting the UV-to-sub-mm SEDs of a sample of IR luminous sources at redshift 2, with an approach based on Bayesian analysis, we have derived a recipe for dust attenuation where the shape of the curve is characterized by a grayer slope than the one of Calzetti~(2000) at all wavelengths. The universality of such a law must be questioned since some variability is expected, as seen in the introduction, as a function of physical and structural parameters of galaxies, galaxy inclination and effective optical depth \citep[e.g.][]{Wild2011b,Chevallard2013,KriekConroy2013}.

For this reason we have selected objects with IR luminosities in the range of LIRGs and ULIRGs, thus similar to those of our z$\sim$2 galaxies, but at lower redshift and we have applied our analysis to these samples. This choice is also motivated by the fact that one of the aims of HELP is to analyze IR selected sources at any redshift.
We focus, in particular, on a sample of (U)LIRGs at z$\sim$0 for which we have a very good data coverage from far-UV to sub-mm and a sample of LIRGs at z$\sim$1 for which we do not have a very good coverage in the rest-frame UV at the given redshift. The first sample, at z$\sim$0, is particularly interesting as it provides us with a class of objects characterized by IR luminosities in the range of our z$\sim$2 (U)LIRGs, but whose nature is likely to be governed by a different physical mechanism (merger-induced starbursts vs cold gas accretion). They have also been shown to lie above the \citet{Meurer1999} relation in the IRX-$\beta$ diagram \citep[see e.g.,][]{Howell2010} which suggests that they can be characterized by grayer attenuation curves than the Calzetti law.     z$\sim$1 LIRGs could also be an interesting case as they dominate the star formation rate density  at z$>$0.8.

The two samples and relative results are presented in the next two sections.

\subsection{Nearby (U)LIRGs}

The sample of nearby (z $<$ 0.083) (U)LIRGs presented here includes 53 LIRGs and 11 ULIRGs spanning Log(L$_{\rm IR}$/L$_{\odot}$) = 11.14-12.57 selected by \citet{U2012} from the flux-limited ($f_{\rm 60 \mu m}  >$ 5.24 Jy) Great Observatories All-sky LIRG Survey (GOALS, \citealt{Armus2009}).
These account for about 30\% of all LIRGs and 50\% of all ULIRGs in GOALS with a median infrared luminosity of log(L$_{\rm IR}$/L$_{\odot}$) = 11.60 and redshift in the range z = 0.012-0.083 (with median z = 0.028 (DL = 119.0 Mpc). Multiwavelength AGN indicators have been used by the authors to select possible AGNs. As here we are focusing on galaxies powered by star formation in order to avoid further degeneracies in the model parameters related to the possibile addition of an AGN component we have restricted our analysis only to the AGN-free (U)LIRGs.

    We first fit the observed SEDs of these galaxies by assuming the mean power-law attenuation curve derived from our previous analysis and therefore characterized by $\delta_{\rm ISM}=-0.48$. We have fixed the value of $\delta_{\rm BC}$ to the  slope of CF00 model, i.e. -0.7 (a different value would not change the results). The quality of the data allows us to obtain good fits to the observed SEDs of nearby GOALS galaxies, with about 78\% of the objects being fitted with a  $\chi^{2}_{\nu} \leqslant 3$.

As a further check we also run our code on the observed SEDs of these local (U)LIRGs by assuming a DBPL-free attenuation curve and thus leaving the slope of the ISM component as free parameter. Interestingly we find an attenuation curve averaged over the entire sample very similar in shape  to the one estimated for our z$\sim$2 (U)LIRGs and thus grayer than Calzetti~(2000) one.
We interpret this result as an evidence of the ability of our dust attenuation recipe to reproduce the properties of nearby merger-induced starburst galaxies too. Indeed, despite the different nature, both local and z$\sim$2 (U)LIRGs are characterized by relatively high optical depths and dust enshrouded star forming environments whose geometrical features can be responsible for similar shapes in the dust attenuation curves.

\subsection{z$\sim$1 LIRGs}

The sample at z$\sim$ 1 considered here includes $\sim$ 130 LIRGs selected from the PEP sample of \citet{Gruppioni2013} in the GOODS-S field with a spectroscopically confirmed redshift in the interval 0.5-1.5 (44 objects in the range 0.76-1.05). Following \citet{LoFaro2013} we selected only sources with 24 $\mu$m flux in the range 0.14-0.45 mJy (same interval as for our z$\sim$2 (U)LIRGs). All the sources benefit from full PACS photometry from 70 to 160 $\mu$m and additional SPIRE data (250, 350 and 500 $\mu$m) retrieved from the HerMES program. The complementary photometry at shorter wavelengths is provided by the MUSIC multiwavelength catalogue by \citet{Santini2009} bringing to a total number of photometric bands of about 19. However, at the redshifts probed by the PEP sample the wavelength coverage does not allow a good sampling of their far-UV colors preventing us from performing a detailed analysis of the UV shape of the dust attenuation curve of these sources.

AGN dominated objects have been excluded by running the spectro-photometric code CIGALE in combination with a reduced \citet{Fritz2006} library of AGN models \citep[see e.g.,][]{Ciesla2015} and computing the AGN fraction contributing to the total IR luminosity of the galaxy.

   We follow a similar  strategy used above for the nearby LIRGs and ULIRGs. We fit the observed SEDs by assuming the mean power-law attenuation curve derived for the z$\sim$2 (U)LIRGs. About 91\% of z$\sim$1 LIRGs are fitted with a $\chi^{2} \leq 3$. We did not try to run the code assuming a DBPL-free attenuation curve since the wavelength coverage  do not allow a fine sampling of the rest-frame UV side of the attenuation curve.

We conclude that a realistic combination of the two main recipes considered in this work should be included in the SED-modelling, especially when dealing with a big database including a large variety of IR bright sources. Within the context of HELP or big legacy projects including multiple surveys, it becomes inefficient to perform SED-fitting including all the possible attenuation laws we do expect. In order to make the process more efficient we are now working to implement in CIGALE a new recipe for  dust attenuation  formalized by a broken power-law with two different slopes for the UV and NIR range, using the V-band wavelength as reference.     In this way,  with one single recipe, we can reproduce both the flattening and steepening of the dust attenuation curve from UV to NIR, including the possibility to have curves which are grayer in the UV but as steep as the Calzetti law in the NIR.

\section{Effect of the assumed dust attenuation curve on the predicted  stellar mass and Star formation rate}
\label{sfr_mass_consequences}

Despite the large efforts devoted to explore the effects that the assumed SFH (e.g., rising, declining or bursty) or wavelength coverage (e.g. availability of IR data) can have on the main physical parameters of galaxies \citep[see e.g.][]{BelldeJong2001, Pforr2012, Michalowski2012, Conroy2013, Buat2014}, there is still a lack of specific discussion in the literature about the effect of the assumed dust attenuation recipes on the derived stellar mass and SFR of galaxies.

\citet{LoFaro2013, LoFaro2015}, by applying radiative transfer models to the sample of dust obscured z$\sim$2 (U)LIRGs analyzed here, observed that the assumed parametrization for the dust attenuation together with the availability of full multiwavelength SED, can also affect the estimates of the stellar mass of galaxies. They found stellar masses systematically higher than those computed with more classical approaches \citep[as e.g. Hyperz by][]{Bolzonella2000}, up to a factor of $\sim$ 6 for the most dramatic cases.

Here we want to explicitly investigate the effect of the assumed shape of the dust attenuation curve on the derived stellar mass and SFR of our z$\sim$2 (U)LIRGs.
\begin{figure*}
\centerline{
\includegraphics[width=9.2cm]{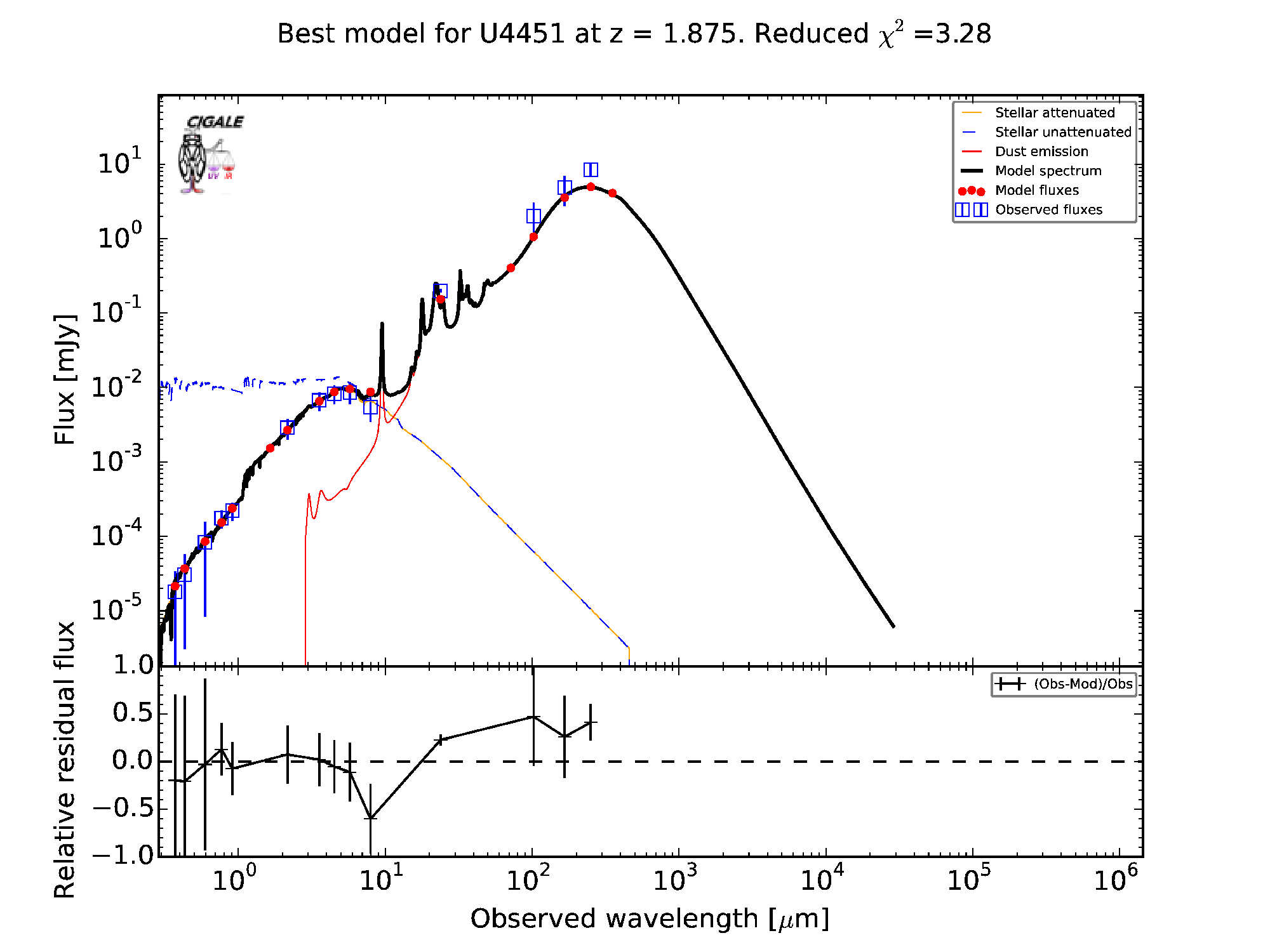}
\includegraphics[width=9.2cm]{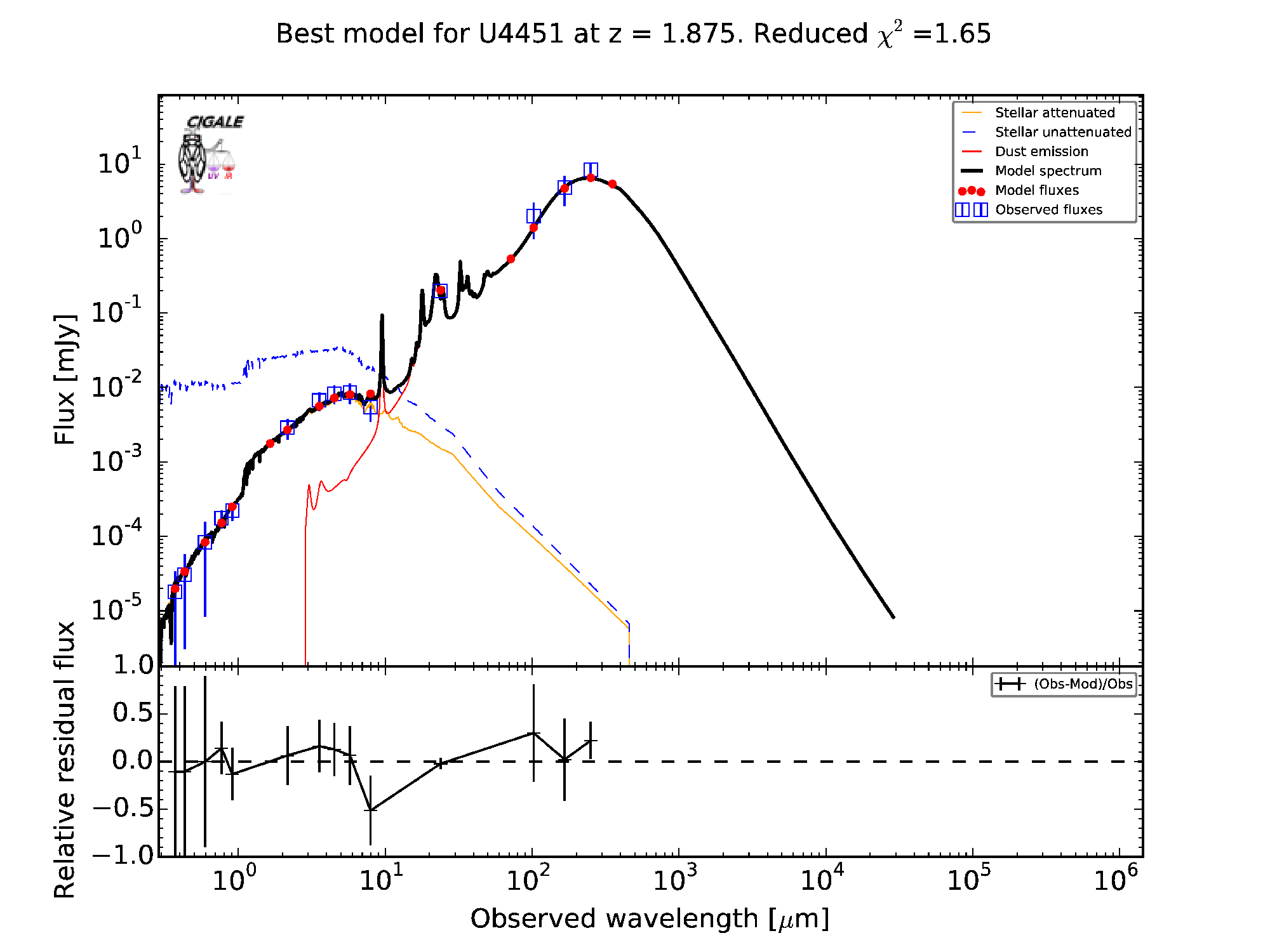}}
\caption{Example of a best-fit SED obtained with CIGALE for a representative object, \textit{U4451}, under the two ``extreme'' assumptions for the dust attenuation law given by \citet{Calzetti2000} (left) and DBPL-free recipe (right). The figure clearly shows the effect on the un-extinguished SED of a grayer attenuation at longer wavelengths (right panel) responsible for an extra attenuation in the NIR.}
\label{sed-nir-effect}
\end{figure*}
\subsection{Stellar mass}
\label{stellar_mass_effect}

To illustrate the effect of a grayer attenuation curve at longer wavelengths on the stellar mass estimate, we show, in Figure~\ref{sed-nir-effect}, the best-fit SEDs of one of our z$\sim$2 (U)LIRGs.

The two fits refer to the object \textit{U4451}, whose properties are representative of the entire galaxy sample and whose attenuation curve is shown in figure~\ref{dust_attenuation_curves}. The best-fit SED obtained under the assumption of a Calzetti~(2000) attenuation law (left side) is compared to that one computed adopting the DBPL-free recipe with $\delta_{\rm ISM}=-0.55$ (right side). Both fit are satisfactory in terms of $\chi^{2}_{\nu}$, with $\chi^{2}_{\nu} (Calzetti~2000)=3.28$ and $\chi^{2}_{\nu} (DBPL-free)=1.65$. The main difference between the two pertains to the rest-frame UV-opt-NIR where the unextinguished stellar SEDs appear quite different. In particular, an extra attenuation at rest-frame NIR wavelengths is visible in the best-fit performed with the DBPL-free law. Differences in the rest-frame UV are also present but they are not as significant as those at longer wavelengths.

For this specific object the logarithm of the stellar mass computed with Calzetti is 10.46 while that one computed assuming the DBPL-free law is 11.13 about 0.6 dex difference.
\begin{figure}
\centering
\includegraphics[width=8.6cm]{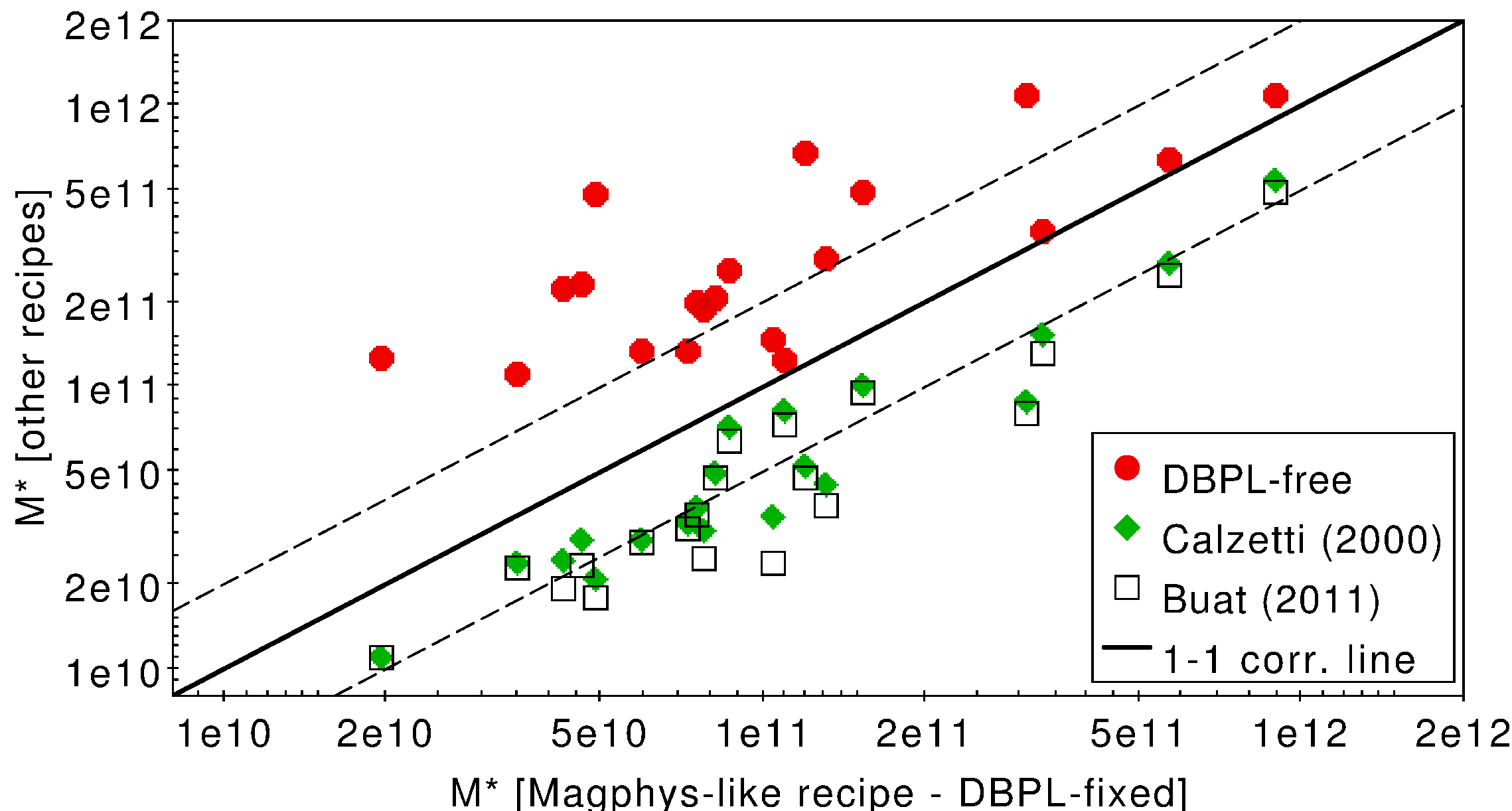}
\caption{Comparison of stellar masses estimated using different assumptions for the dust attenuation curve. MAGPHYS based M$_{\star}$ on the x-axis are compared to the stellar masses computed with all the other recipes on the y-axis. The solid black line represents the 1-1 correlation line while the dashed lines highlight the region within a factor of 2 from the 1-1 corr. line. The results from the DBPL-free recipes are shown as filled red circles, those from \citet{Calzetti2000} as filled green rhombus while \citet{Buat2011} modified Calzetti recipe as open black squares.}
\label{mstar-diff}
\end{figure}

Figure~\ref{mstar-diff} shows the global effect of the assumed attenuation curve on the estimated stellar mass of all the (U)LIRGs in our sample for all the recipes considered. The MAGPHYS (DBPL with $\delta_{\rm ISM}$=-0.7 and $\delta_{\rm BC}$=-1.3) based stellar masses on the x-axis are compared to the stellar masses computed with all the other recipes on the y-axis.
The power law recipes adopted in the DBPL-free and MAGHYS-like models provide, on average, larger stellar masses compared to Calzetti-like recipes  because they consider flatter  attenuation curves in the NIR. The difference between the stellar mass estimates is higher for the DBPL-free configuration which allows even grayer slopes for the $\delta_{\rm ISM}$ than the value of MAGPHYS ($\delta_{\rm ISM}=$-0.7) and it can reach up to a factor of $\sim$ 10 for the most dust obscured and IR luminous objects (red filled circles in Fig.~\ref{mstar-diff}). The observed trend is in agreement with the results from \citet{LoFaro2013} and \citet{Mitchell2013}.
A similar comparison for the two samples of IR luminous objects at z~$\sim$0 and z$\sim$1 is given in the Appendix.

\subsection{Star formation rate}
\label{SFR_effect}

\begin{figure}
\centerline{
\includegraphics[width=9.cm]{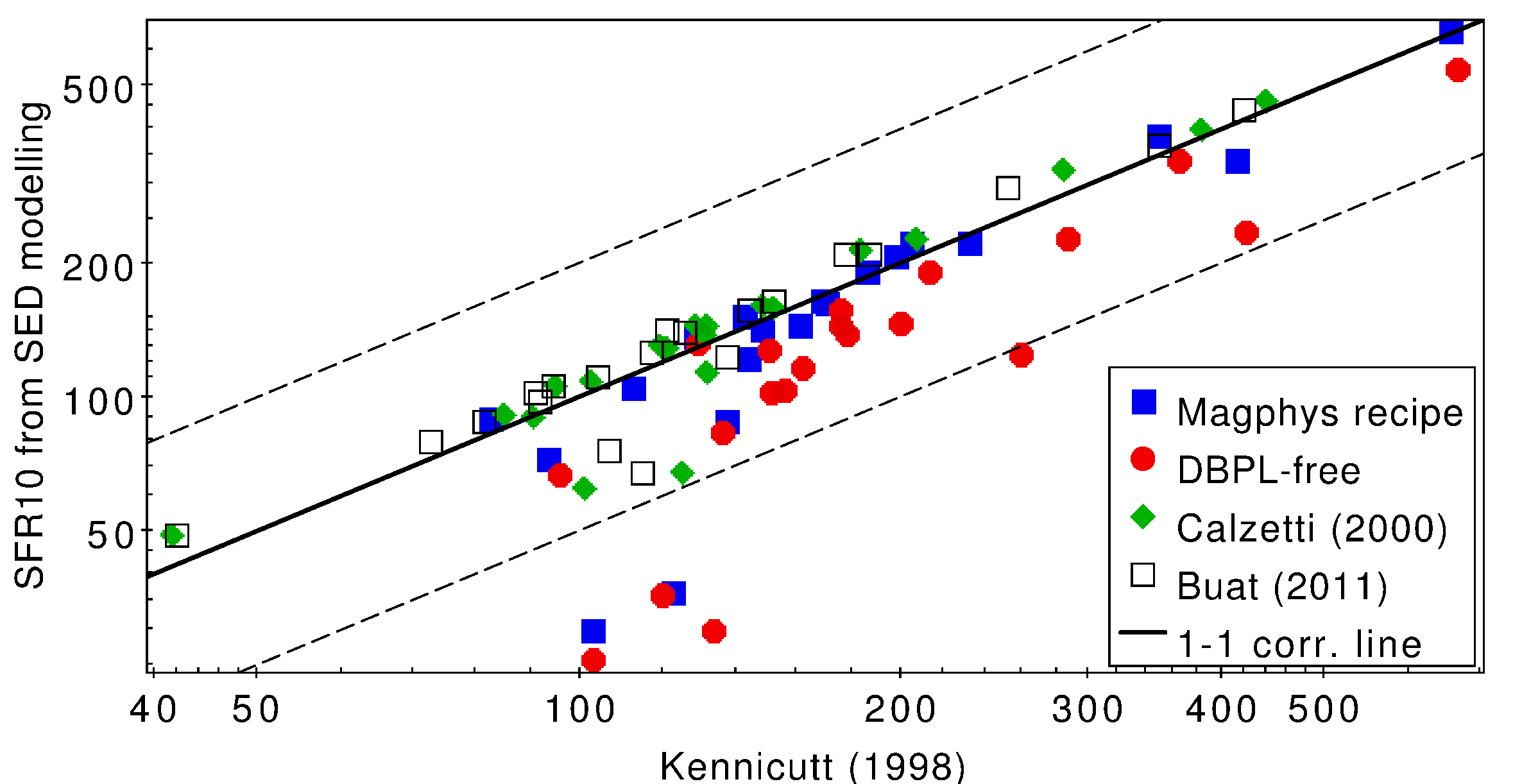}}
\caption{Comparison of the SFRs derived, for our galaxies, from SED-fitting and averaged over the last 10 Myr with the classical \citet{Kennicutt1998} calibration. The SFRs derived from SED modelling are color coded as as function of the different assumed dust attenuation curves, MAGPHYS-like as filled blue squares, the DBPL-free model as filled red circles, the \citet{Calzetti2000} recipe as filled green rhombus and the modified Calzetti recipe by \citet{Buat2011} as open black squares. The black solid line highlights the 1-1 correlation line while the two dashed lines define the region within a factor of 2 from the 1-1 correlation line.}
\label{sfr-diff}
\end{figure}
The differences observed in the UV range among the different recipes considered translate into different stellar population ages and total FUV attenuation which can affect the derived SFR of our high-$z$ (U)LIRGs. In Figure~\ref{sfr-diff} we compare the well-known and widely used in literature \citet{Kennicutt1998} calibration (x-axis) to the SFRs derived from SED-fitting and averaged over the last 10 Myrs. The latter are color coded as a function of the assumed dust attenuation law.

    The situation appears less dramatic than that one observed for the stellar mass. There is an overall good agreement among the different SFR estimates well within a factor of two, which is the typical uncertainty expected when comparing different recipes in the context of SED-modelling \citep[see e.g.,][]{LonghettiSaracco2009}. Despite the global agreement there is a tendency for the DBPL-free estimates (red circles) to be lower on average than both those derived assuming Calzetti-like recipes and those computed using the \citet{Kennicutt1998} calibration due to the progressive flattening of the dust attenuation curve in the UV. We find the difference between the SFRs estimated with Calzetti and power-law recipes to be larger for larger differences of the measured FUV total attenuation.

For three objects the discrepancy between the SFR estimated with our DBPL-free model (red circles in figure~\ref{sfr-diff}) and that one derived from the Kennicutt calibration is larger than a factor of two.  \textit{U5795} at the bottom right of figure~\ref{sfr-diff} is the object showing the flattest attenuation curve ($\delta_{\rm ISM}=-0.1$) and identified as very badly fitted (section 4.3). It is probably indicative of our inability to constrain its dust properties with the available data.
For the other two sources falling below the bottom dashed line in figure~\ref{sfr-diff}, namely \textit{U4367} and \textit{U5050} our DBPL-free  fit is very good ($\chi^{2}_{\nu}\simeq 1.3$), both the DBPL-free (red circles) and MAGPHYS (blue squares) recipe provide SFRs which differ from those derived from the Kennicutt calibration by a factor larger than two (the consistency between the results found with both recipes is expected given the similarity of the slope of the attenuation curve for the ISM component).  For  these two objects even the SFRs estimated by assuming a Calzetti-like dust law (green rhombus/black open squares), lie at the limits of the dashed line delimiting the region within a factor of 2 from the 1-1 corr. line, when compared to the Kennicutt calibration. Looking at the observed SEDs we noticed that they present the most  prominent   bump in the rest-frame optical-NIR among all the galaxies in the sample and also older ages than the others. Their IR SEDs do not present any deviant property compared to the other galaxy IR SEDs.
We thus focus the exploration of the possible reasons at the origin of the observed discrepancy in the SFR estimates on the star formation history of these galaxies.

\begin{figure*}
\centerline{
\includegraphics[width=9.cm]{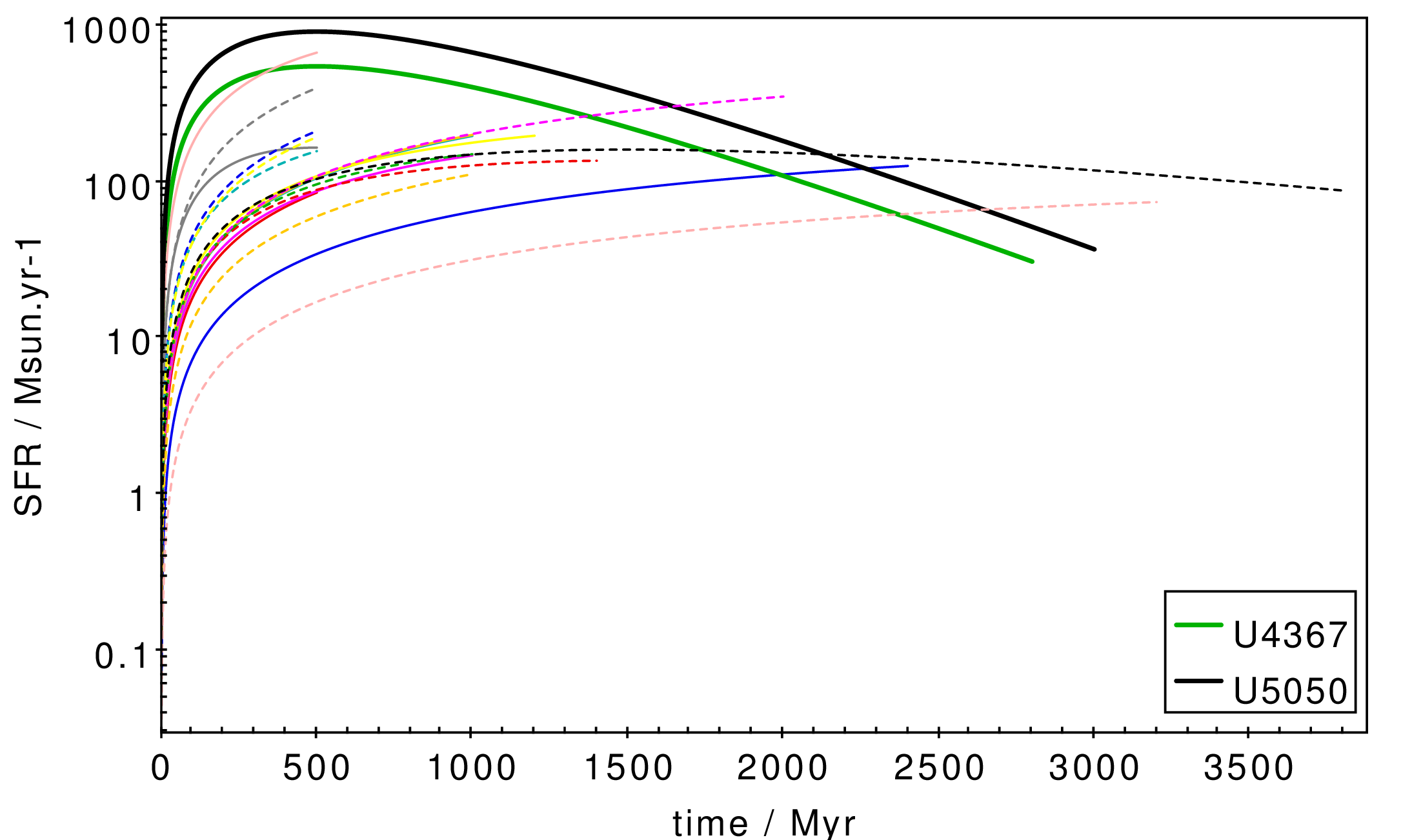}
\includegraphics[width=9.1cm]{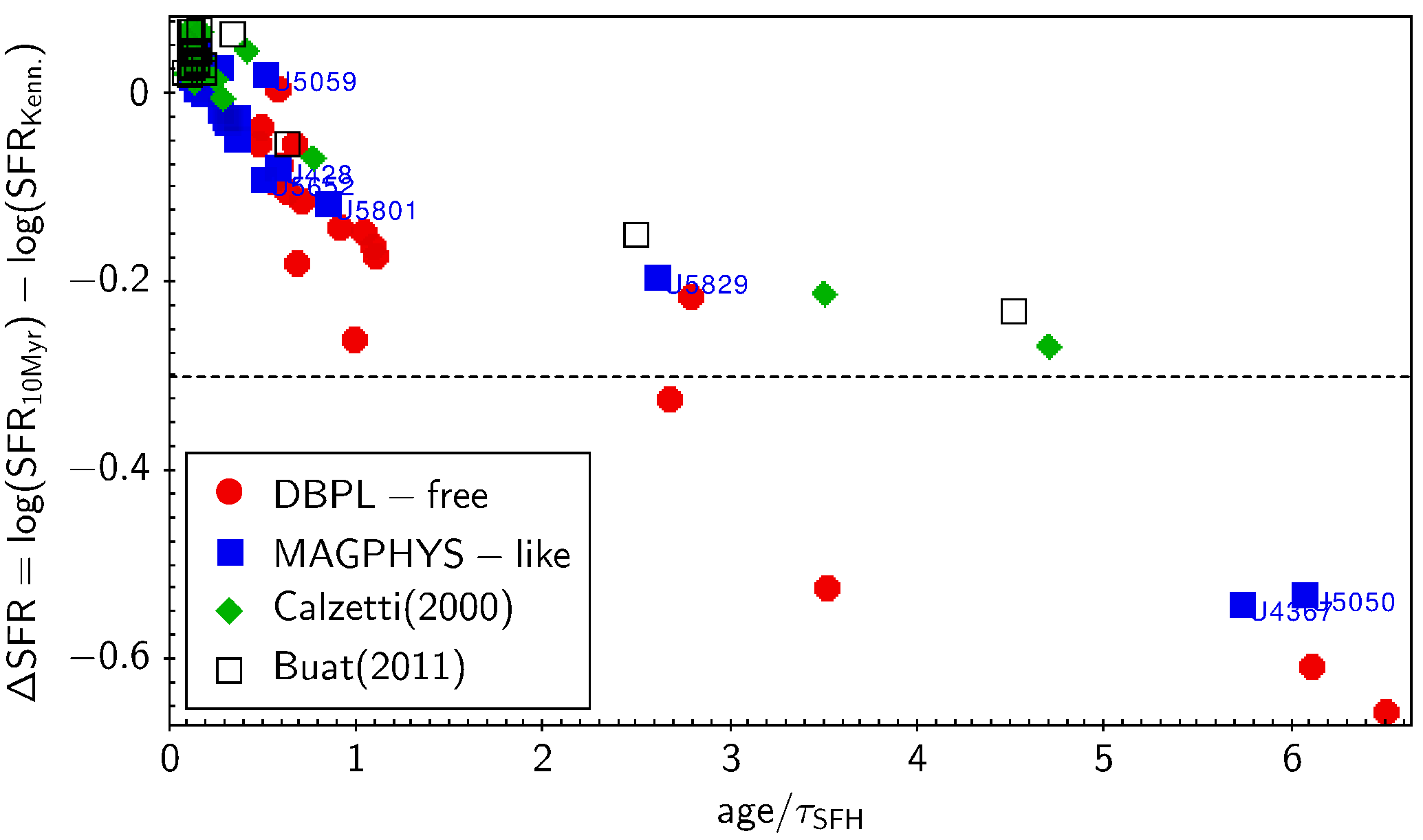}}
\caption{Left: best-fit SFHs of our z$\sim$2 (U)LIRGs as derived from our SED-fitting procedure under the assumption of double power-law attenuation curve, corresponding to the filled blue squares of figure~\ref{sfr-diff}. The two thick green and black lines represent the two objects showing deviant behavior in terms of SFR in figure~\ref{sfr-diff}. Their SFHs are consistent with them being observed well after the peak of SF. Right: Difference $\Delta SFR=Log(SFR_{10})-Log(SFR_{\rm Kenn})$ of the SFRs derived from SED-fitting and Kennicutt calibration plotted as a function of the ``evolutionary parameter'' $age/\tau_{\rm SFH}$ indicating how far from the peak we are observing the galaxy (fig.~\ref{SFH_ulirgs} - right panel). The largest discrepancies in SFR, above a factor of $\sim$2-2.5, are observed in correspondence of values of the ``evolutionary parameter'' above 2.5.}
\label{SFH_ulirgs}
\end{figure*}

Figure~\ref{SFH_ulirgs} (left panel) shows, for the DBPL recipe, the best-fit delayed-$\tau$ SFHs obtained through our SED-fitting procedure for each of the z$\sim$ 2 (U)LIRGs in our sample. The SFHs are dropped in correspondence of the age of the galaxy as derived from our analysis. The thick green and black lines represent, respectively, the delay-$\tau$ SFH of \textit{U4367} and \textit{U5050}. Differently from the other galaxies, these are the only two objects observed not at the peak of their star forming activity. We can use the ``evolutionary parameter'', defined as the ratio $age/\tau_{\rm SFH}$, to quantify how far from the peak a galaxy is observed. A clear anti-correlation is measured between the difference in the SFRs estimated from SED-fitting and Kennicutt calibration, $\Delta SFR=Log(SFR_{\rm 10 Myr})-Log(SFR_{\rm Kenn})$, and the evolutionary parameter (Figure~\ref{SFH_ulirgs} - right panel). The plot shows that the largest discrepancies in SFR, above a factor of $\sim$2-2.5 (as for \textit{U4367} and \textit{U5050}), are reached for values of the ``evolutionary parameter'' above 2.5.
The plot also shows that for the Calzetti-like recipes the SFHs are mostly consistent with the galaxies being observed close to the peak. We interpret the fact to be observed well after the peak of star formation as the most likely cause of the deviant behaviour in terms of SFR. This is consistent with the prominent  bump     in the NIR characterizing the observed SED of the two objects mentioned above. In these cases we may expect the Kennicutt calibration to overestimate the SFR due to a non-negligible contribution of relatively evolved stellar populations to the infrared luminosity \citep[see e.g. Fig.~4 in][]{LoFaro2013}

A similar comparison between the SFRs derived from SED-fitting and those based on the \citet{Kennicutt1998} calibration is performed for the two samples of IR luminous galaxies at lower redshift discussed in Section~\ref{low_redshift_checks} and  is presented in the Appendix.

\section{Summary \& Conclusions}

We have investigated the shape of the dust attenuation curve of a sample of dust obscured IR luminous  galaxies at z$\sim$2 by fitting their UV-to-sub-mm SEDs by means of the physically-motivated code CIGALE including energy balance. The high flexibility offered by CIGALE allowed us to implement several different recipes for the dust attenuation, both well-known and new. These include the classical \citet{Calzetti2000} law, a modified version of this curve which includes a UV-bump and steeper slope in the UV \citep[][]{Buat2011}, and our newly implemented two component (birth clouds and diffuse ISM) power-law recipe (DBPL-free) based on the formalism of \citet{CharlotFall00}. In our DBPL-free model, used here as reference, the UV-to-NIR shape of the dust attenuation curve is treated as a free parameter and implicitly includes both the \citet{CharlotFall00} and MAGPHYS-like prescriptions.

To focus our analysis on the characterization of the dust attenuation properties of these galaxies we assumed the same configuration for the input SFH, stellar library and IR emission, for all the different dust attenuation recipes considered.
We explored the different recipes for dust attenuation and compared the results to those derived in previous works for the same galaxy sample with radiative transfer computations.

The comparison revealed to be  successful ifor our DBPL-free reference model whose attenuation curves appear to be in very good agreement with those based on RT models     for $\sim 75 \%$ of the objects and grayer at all wavelengths than the Calzetti law. A grayer attenuation law also explains the position of our galaxies in the IRX-$\beta$ diagram, above the \citet{Meurer1999} relation for local, UV selected starbursts.

Our analysis thus shows that for IR selected and luminous galaxies we do expect a global flattening of the dust attenuation curve at all wavelengths from UV-to-NIR. In agreement with radiative transfer models we observe grayer slopes for the ISM component of the medium at increasing optical depths. This flattening can be explained by mixed star-dust geometries including clumping of both components.

We have checked our recipe for the dust attenuation by applying our analysis to two samples of IR bright sources with IR luminosities similar to those of our z$\sim$2 galaxies. These include a sample of $\sim$ 64 nearby (U)LIRGs for which a wealth of UV-to-sub-mm data is available and $\sim$130 LIRGs at z$\sim$1 with a good spectral coverage except for the rest-frame UV.

We propose a  flexible recipe for the dust attenuation curve formalized by a broken power-law with two different slopes for the UV and NIR range. This recipe would allow us to reproduce both the flattening and steepening of the dust attenuation curve from UV to NIR, including the possibility to have curves which are grayer in the UV but as steep as Calzetti in the NIR, mimicking, implicitly, the effect of different geometrical configurations.

We finally investigated the effect of these grayer attenuation curves on the derived main physical properties of galaxies and found, in agreement with \citet{LoFaro2013} and \citet{Mitchell2013}, that they can strongly affect the estimate of the stellar mass of galaxies up to a factor $\lesssim$ 10 for the most extreme cases (with the median factor being $\sim$1.4). Larger attenuations at longer wavelengths translate into larger stellar masses compared to classical Calzetti-like recipes. The SFR appears to be less affected by these variations on the specific shape of the dust attenuation curve. In 2 cases the SFH is suggested to play a role in the SFR determination with a lower SFR obtained from SED-fitting with respect to that one based on the Kennicutt~1998 calibration, for those galaxies observed far from the peak of SF.

\section*{Acknowledgments}

The project has received funding  from the European Union Seventh Framework Programme FP7/2007-2013/ under grant agreement n$^{\circ}$ 60725. B.LF and Y.R acknowledge support from this programme. This publication reflects only the authors view and the European Union is not responsible for any use that may be made of the information contained therein.




\bibliographystyle{mnras}
\bibliography{Bibliography} 
\appendix
\section{Effect of the assumed attenuation curve on the estimated M$_{\star}$ and SFR of nearby and z$\sim$1 (U)LIRGs}
We present here the results concerning the global effect of our agreed prescription for the dust attenuation on the derived stellar mass and SFR also for the two samples at lower redshift presented in Section~\ref{low_redshift_checks}. As we already discussed the differential effect among all the different recipes for our reference sample of z$\sim$2 (U)LIRGs, we restrict here the analysis to the two main formalisms considered under the \citet{Calzetti2000} and DBPL-free configuration. The latter implicitly includes the standard CF00 model, the MAGPHYS-like recipe and our ``mean'' attenuation curve derived from the analysis of the z$\sim$2 (U)LIRGs. The results are shown in Figure~\ref{mstar_sfr_lowz} with the nearby GOALS galaxies represented as red circles and the z$\sim$1 LIRGs as blue squares. As for the z$\sim$2 (U)LIRGs also in this case we find the stellar masses estimated under the Calzetti recipe to be systematically lower compared to those derived with DBPL attenuation law due to the flattening at longer wavelengths. Interestingly the effect is stronger for the local LIRGs showing the larger NIR attenuations ($A_{\rm NIR}$). The lower attenuations at longer wavelengths obtained for the z$\sim$1 LIRGs under the DBPL recipe, together with a slope of the attenuation curve being consistent in the UV with the Calzetti recipe, provide stellar masses which are systematically lower than those computed with Calzetti but mostly within a factor of 2 from the 1-1 correlation line (black solid line in the figure).
\begin{figure}
\includegraphics[width=8 cm]{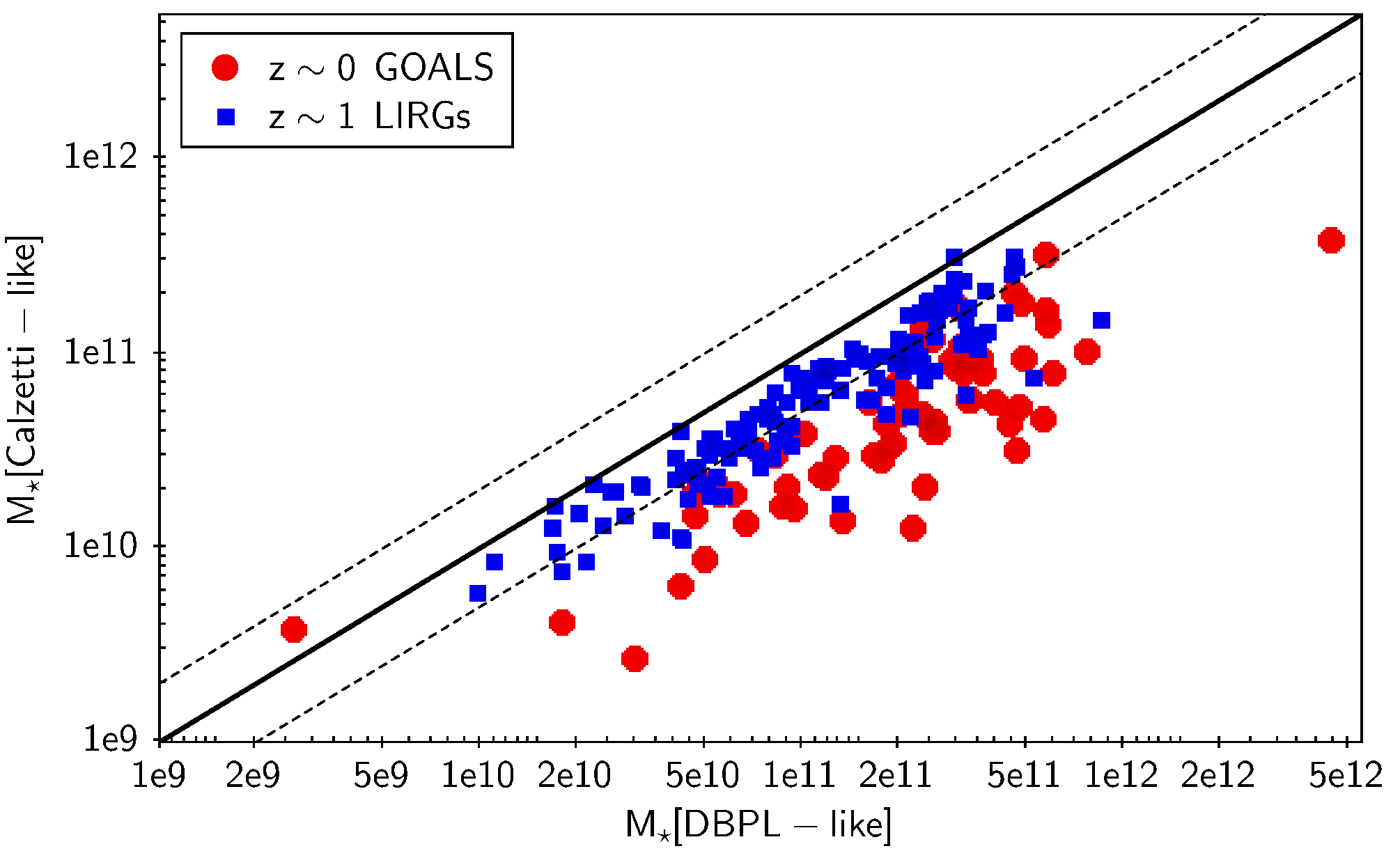}
\includegraphics[width=8 cm]{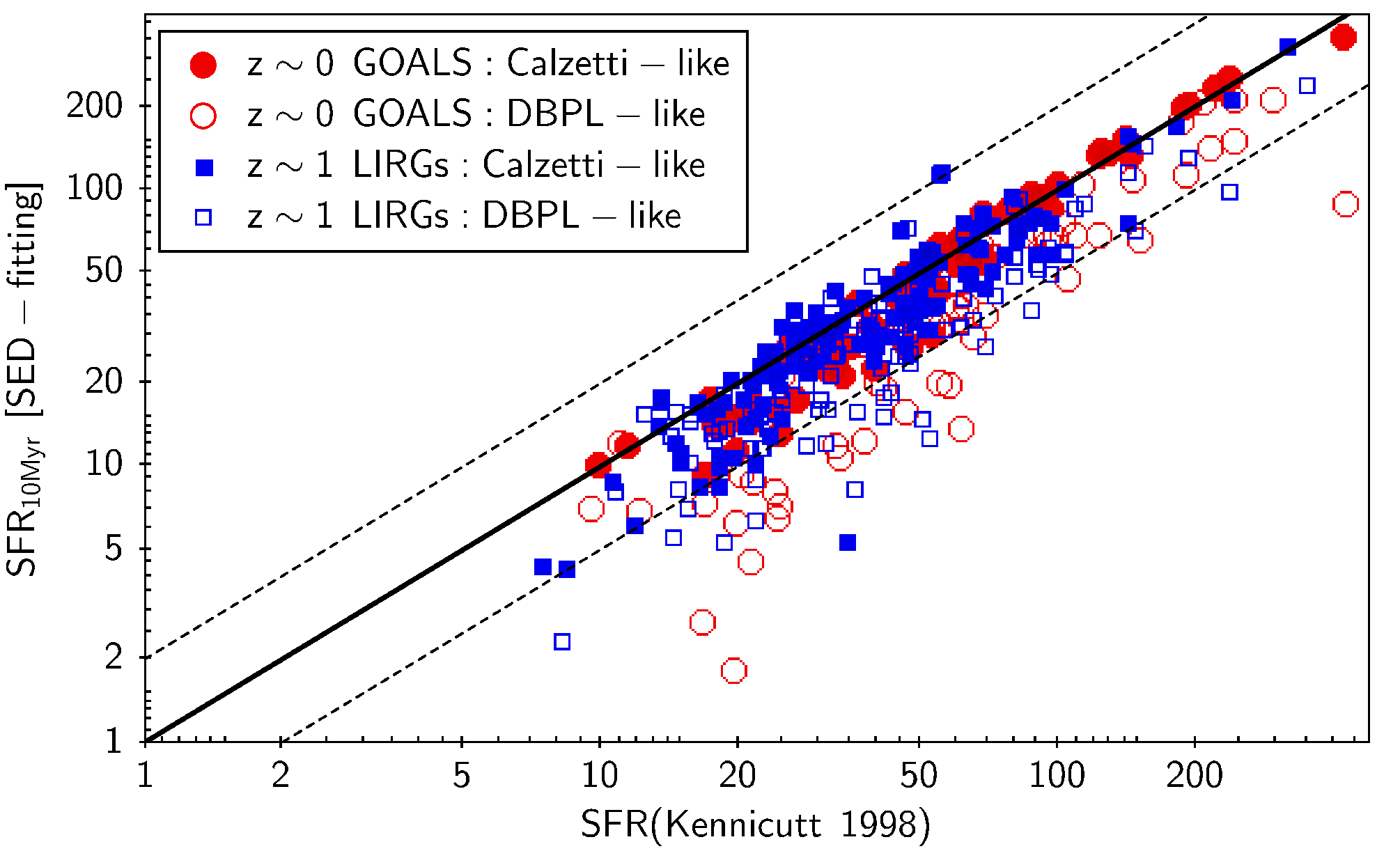}
\caption{Top: the stellar masses derived under the DBPL configuration (x-axis) are compared to those derived by assuming \citet{Calzetti2000} attenuation law (y-axis). Red filled circles are for the nearby GOALS galaxies while filled blue squares are for the z$\sim$1 LIRGs, both presented in Section~\ref{low_redshift_checks}. The black solid line represents the 1-1 correlation line while the two dashed line define the area within a factor of 2 from the 1-1 corr. line. Bottom: the SFRs estimated by using the \citet{Kennicutt1998} calibration (x-axis) are compared to those derived from SED-fitting (y-axis) under the two main assumptions for the dust attenuation of DBPL-like (open symbols) and Calzetti-like (filled symbols). }
\label{mstar_sfr_lowz}
\end{figure}
A better agreement is found when comparing the SFRs derived from SED-fitting with those based on the \citet{Kennicutt1998} calibration and it is shown in Figure~\ref{mstar_sfr_lowz} (right panel). In this case both nearby (red circles) and z$\sim$1 (blue squares) LIRGs present similar behaviour. On average there is a very good agreement between the SFRs computed under the Calzetti-recipe and those derived from the Kennicutt calibration while those based on the DBPL configuration appear to be lower on average but mostly within a factor of 2 from the 1-1 correlation line (black solid line in the figure). The fraction of objects falling below the above mentioned area are those showing the most prominent flattening in the UV.


\label{lastpage}

\end{document}